# Keyword Targeting Optimization in Sponsored Search Advertising: Combining Selection and Matching

Huiran Li[1] and Yanwu Yang[2]

[1]School of Business Administration and Customs Affair, Shanghai Customs College, Shanghai 201204, China

[2]School of Management, Huazhong University of Science and Technology, Wuhan 430074, China

{lihuiran.isec, yangyanwu.isec}@gmail.com

***Abstract:*** In sponsored search advertising (SSA), advertisers need to select keywords and determine matching types for selected keywords simultaneously, i.e., keyword targeting. An optimal keyword targeting strategy guarantees reaching the right population effectively. This paper aims to address the keyword targeting problem, which is a challenging task because of the incomplete information of historical advertising performance indices and the high uncertainty in SSA environments. First, we construct a data distribution estimation model and apply a Markov Chain Monte Carlo method to make inference about unobserved indices (i.e., impression and click-through rate) over three keyword matching types (i.e., broad, phrase and exact). Second, we formulate a stochastic keyword targeting model (BB-KSM) combining operations of keyword selection and keyword matching to maximize the expected profit under the chance constraint of the budget, and develop a branch-and-bound algorithm incorporating a stochastic simulation process for our keyword targeting model. Finally, based on a realworld dataset collected from field reports and logs of past SSA campaigns, computational experiments are conducted to evaluate the performance of our keyword targeting strategy. Experimental results show that, (a) BB-KSM outperforms seven baselines in terms of profit; (b) BB-KSM shows its superiority as the budget increases, especially in situations with more keywords and keyword combinations; (c) the proposed data distribution estimation approach can effectively address the problem of incomplete performance indices over the three matching types and in turn significantly promotes the performance of keyword targeting decisions. This research makes important contributions to the SSA literature and the results offer critical insights into keyword management for SSA advertisers.

# 1. Introduction

Sponsored search advertising (SSA) has become one of the most indispensable digital media channels. In the United States, SSA spending is projected to reach $171,641 million in 2021 (Statista, 2021; IAB, 2021). SSA is a prosperous market with three types of players, i.e., search users, advertisers and search engines, where keywords serve as a bond tying all three together (Yang et al., 2019). In SSA, advertisers have to select an appropriate set of keywords and determine matching types for selected keywords simultaneously. This process is called keyword targeting (Yang et al., 2019). Keyword targeting controls the aggressive and restrictive degree to which consumers' searches trigger sponsored search auctions, and helps advertisers better fit their promoted products to search engines (Kiritchenko and Jiline, 2008)[1]. Well-targeted keywords will guarantee that the right advertisements are delivered to the right consumers (Yang et al., 2017). Therefore, it is critical for advertisers to effectively make keyword targeting decisions for their SSA campaigns.

In the literature on SSA, plenty of research efforts have been invested in formulating keyword selection models and developing corresponding solution algorithms (e.g., Rusmevichientong and Williamson, 2006; Kiritchenko and Jiline, 2008; Zhang et al., 2014), analyzing branded and competitor's keywords (e.g., Desai et al., 2014), and examining keywords' performance (e.g., Lu and Yang, 2017). In another independent research stream, keyword matching has been studied extensively in recent years from various aspects: identifying high-quality keywords (Radlinski et al., 2008; Gupta et al., 2009; Grbovic et al., 2016), profiling advertising metrics over matching types (e.g., Ramaboa and Fish, 2018), and bidding optimization over broad match (e.g., Singh and Roychowdhury, 2008; Even Dar et al., 2009; Amaldoss et al., 2016). Operationally, SSA systems require advertisers to select a set of keywords and determine how these keywords will be matched to search queries (i.e., broad match, phrase match or exact match) at the same time (Du et al., 2017). In effect, decisions over multiple keywords might be interdependent to each other (Yang et al., 2019). Moreover, advertisers need to make advertising decisions in realtime due to the ever-changing nature of SSA environments (Yang et al., 2022). Conceptually, joint optimization for several related advertising decisions can significantly improve the performance of SSA campaigns (Yang et al., 2012; Zhang et al., 2012; Nuara et al., 2022). Thus, from both operational and theoretical perspectives, it's of necessity to address keyword selection and keyword matching problems in an integrated way (i.e., keyword targeting), in order to help advertisers effectively reach the targeted population. To the best of our knowledge, there is no study on keyword targeting for search advertising campaigns. This paper aims to fill this crucial gap in the literature.

---

[1] https://support.google.com/google-ads/answer/7478529?hl=en

In SSA, advertisers have to face many challenges while making keyword targeting decisions. First, advertisers have no complete performance information over the three keyword matching types for each keyword. In practice, keyword-level historical records only contain performance indices (e.g., impressions and click-through rate) for a certain matching type chosen in past advertising campaigns. When making keyword selection decisions, advertisers have to take the uniform assumption about performance indices over the three keyword matching types, which certainly results in suboptimal solutions because advertising performance indices are systematically different over the three keyword matching types (Ramaboa and Fish, 2018; Yang et al., 2021a). Second, the SSA environment is highly uncertain (Yang et al., 2013; Li and Yang, 2020). In such an uncertain market, advertisers must make keyword targeting decisions prior to the realization of values for keyword performance indices (Amaldoss et al., 2016). In addition, advertisers mostly have limited budgets for SSA campaigns (Yang et al., 2015). That is, advertisers need to select appropriate keywords and determine matching types with consideration of their budget constraints.

The objective of this research is to address the keyword targeting problem in the SSA context. First, we construct a data distribution estimation model for keyword performance indices over the three keyword matching types. It is supposed that performance indices follow the multivariate normal distribution over the three matching types. The Markov Chain Monte Carlo method is applied to make inference about unobserved performance indices. Second, we formulate a stochastic keyword targeting model (BB-KSM for short) to maximize the expected profit under the chance constraint of the budget, and develop a branch-and-bound algorithm incorporating a stochastic simulation process for our keyword targeting model. Finally, using a realworld dataset collected from field reports and logs of past SSA campaign, a series of computational experiments are conducted to evaluate the performance of our keyword targeting strategy against seven baselines.

Experimental results show that (a) BB-KSM performs better than seven baselines in terms of profit; (b) BB-KSM increasingly shows its superiority as the budget increases, especially in situations with much more keywords and keyword combinations; (c) our data distribution estimation approach can effectively address the problem of incomplete information of performance indices over keyword matching types and in turn significantly promote the performance of keyword targeting decisions. In keyword targeting, strategies with mixed keyword matching enriches keyword portfolios and increases the expected profit, compared with those with a single matching type. BB-KSM can help advertisers find more high-profit yet low-cost keywords with appropriate matching types by searching the keyword targeting space for the global optimum. This research contributes to the SSA literature and offers critical implications for SSA advertisers.

The remainder of this paper is structured as follows. Section 2 provides a brief literature review. In Section 3, we build a data distribution estimation model and a stochastic keyword targeting model in SSA, and develop algorithms for our models. In Section 4, we conduct computational experiments and report results. Finally, we conclude this research in Section 5.

## 2. Related Work

In the SSA field, plentiful research efforts have been made to explore search auction mechanism design (e.g., Huang and Kauffman, 2011; Yang et al., 2020) and search user behavior analysis (e.g., Lo et al., 2014; Vragov et al., 2019; Lian et al. 2021), empirical analysis of performance indices (e.g., Yang et al., 2018; Jeziorski and Moorthy, 2018; Schultz, 2020; Yang and Zhai, 2022), and advertising decisions including bidding optimization (e.g., Küçükaydin et al., 2020; Kim et al., 2021), budget optimization (e.g., Yang et al., 2012; Yang and Xiong, 2020; Avadhanula et al., 2021; Yang et al., 2021b), and keyword optimization (e.g., Qiao et al., 2017; Nie et al., 2019; Scholz et al., 2019; Song et al., 2021; Zhang et al., 2021). This study focuses on one particular type of keyword decisions, i.e., keyword targeting, which draws from two research streams, namely keyword selection and keyword matching.

### 2.1 Keyword Selection

Keyword selection is the basis for the effectiveness of SSA campaigns (Szymanski and Lininski, 2018). Researchers have addressed the keyword selection problem through semantic mapping (Kiritchenko and Jiline, 2008; Arroyo-Cañada and Gil-Lafuente, 2019; Nagpal and Petersen 2020) and optimization techniques (Rusmevichientong and Williamson, 2006; Zhang et al., 2014; Yang et al. 2019; Symitsi et al., 2022). Based on the feature selection paradigm, Kiritchenko and Jiline (2008) analyzed the historical performance of individual words and phrases generated from users'

queries, selected the most promising keywords extended with highly predictive (positive and negative) words to maximize the profit, and showed that their approach can obtain high-quality keywords and discover specific combinations of keywords. Arroyo-Cañada and Gil-Lafuente (2019) developed a TOPSIS (Technique for order of preference by similarity to ideal solution)-based method sorting keywords according to their distance to the positive and negative ideal solutions, and proved that the proposed method was effective in increasing brand awareness and traffic volumes. Nagpal and Petersen (2020) constructed a conceptual framework to identify profitable keywords by controlling the endogeneity of competition to measure keyword relevance.

Keyword selection decisions can also be taken as optimization problems. Rusmevichientong and Williamson (2006) identified a profitable set of keywords by sorting keywords in the decreasing order of profit-to-cost ratio, and formulated keyword selection as a multi-armed bandit problem, taking into account the uncertainty of click-through rate. Under the budget constraint, Zhang et al. (2014) took the keyword selection as a mixed integer programming problem with objectives of maximizing the profit and the relevance of selected keywords and minimizing the competitiveness of these keywords, presented a sequential quadratic programming solver, and showed that their method can help increase the revenue for advertisers and search engines. With consideration of the entire lifecycle of SSA campaigns, Yang et al. (2019) developed a multilevel keyword optimization framework to handle various keyword decisions (e.g., keyword generation, keyword selection and keyword assignment), and showed that the proposed framework could reach the optimum in a steady way. Based on Markowitz portfolio theory, Symitsi et al. (2022) utilized the risk-adjusted performance to construct diversified keyword portfolios and decide which keywords to be selected and how much to spend on selected keywords.

Another research branch focuses on empirical analysis for keyword selection, e.g., analyzing the competitor's keywords (Desai et al., 2014) and keyword attributes (Lu and Yang, 2017). In order to understand strategic benefits and costs of selecting keywords about advertisers' own brand names and their competitors' brand names, Desai et al. (2014) modeled the effectiveness of SSA campaigns depending on whether competitors' advertisements are presented on the same results page, and found that selecting keywords about their own brand names can preclude their competitors from buying the same keywords, while if both advertisers and their competitors select their brand names, a prisoner's dilemma may be created to hurt both of their profits. Lu and Yang (2017) regarded each keyword as a market and developed a structural model to empirically investigate the spillover effects in advertisers' keyword market entry decisions, i.e., the probability that advertisers use a keyword is affected by their competitors' keyword decisions, and demonstrated that the keyword-specific competition can improve search engine's revenue by around 5.7 percent.

**2.2 Keyword Matching**

Our work also builds on the literature on keyword matching in SSA. Choosing a suitable keyword matching type for each keyword is essential to the success of advertising campaigns (Ghose and Yang, 2009; Li et al., 2016; Ramaboa and Fish, 2018). In general, there are three keyword matching types (i.e., broad match, phrase match and exact match) to choose in SSA, as shown in Table 1.

**Table 1. Three Keyword Matching Types**

| Matching Type | Definition | Example Keywords | Matching Keywords |
|---|---|---|---|
| Broad match[2] | searches that include misspellings, synonyms, related searches, and other relevant variations | internet advertising | Adwords |
| Phrase match | searches that match a phrase, or close variations of that phrase, which may include additional words before or after | internet advertising | internet advertising in China |
| Exact match | searches that match the exact term or are close variations of that exact term | internet advertising | internet advertising |

Different matching types lead to different advertising performance for keywords (Ramaboa and Fish, 2018). In the field of keyword matching, most research focuses on broad match. On one hand, broad match helps advertisers extend keywords that match users' intent expressed by their queries (Radlinski et al., 2008). On the other hand, broad match helps search engines engage more competitors in keyword auctions through broader targeting, thus increasing their revenues (Levin and Milgrom, 2010).

SSA allows advertisers to target a large amount of queries but only bids a few keywords with the support of broad match. Existing literature in this area has explored broad match from both direct and indirect perspectives in SSA. The first research stream studied broad match directly by investigating effective broad match mapping mechanisms to help advertisers identify similar

---

[2] In Google Ads, there exists another type of matching option, "Broad match modifier", which is similar to broad match, except that the broad match modifier option only shows ads in searches that includes the words with a plus sign. We don't distinguish broad match and broad match modifier in this research.

keywords and thus increasing their advertising reach and reducing their campaign management burden (Jones et al., 2006; Radlinski et al., 2008; Gupta et al., 2009; Grbovic et al. 2016). Some approaches require offline training based on human-labeled relevance judgments and use similarity characteristics to identify associated keywords (Jones et al., 2006; Radlinski et al., 2008). Other approaches don't assume the availability of human supervision. For example, Gupta et al. (2009) proposed a machine learning approach utilizing implicit feedback from click-through logs to identify high-quality broad match mappings and estimate click-through rate; Grbovic et al. (2016) proposed a matching strategy based on semantic embeddings to learn queries and ads from user's search sessions, and showed that their strategy had a good performance in terms of relevance, coverage, and the growth of profit.

The second research stream studied broad match indirectly by addressing optimization problems in the context of broad match (Singh and Roychowdhury, 2008; Even Dar et al., 2009; Mahdian and Wang, 2009; Amaldoss et al., 2016). Singh and Roychowdhury (2008) provided a framework to explore economic outcomes of broad match, and observed that if the quality of broad match is good, the auctioneers (i.e., search engines) can always improve their revenue by judiciously using broad match. Mahdian and Wang (2009) developed a clustering-based bidding language to reduce the bidding complexity and avoid the negative economic effect of broad match. Researchers at Google studied bid optimization over broad match (Even Dar et al., 2009), and developed a linear programming based polynomial-time algorithm to obtain the optimal profit. Focusing on equilibrium analysis of broad match, Amaldoss et al. (2016) examined advertisers' strategies and profits over broad match using a game-theoretic model, and showed that there exists an accuracy degree of broad match where advertisers are willing to choose broad match and search engines get the highest profit.

Recent studies have the role of keyword matching types in achieving the advertising effectiveness. Ramaboa and Fish (2018) analyzed the influence of various SSA metrics (i.e., the length, click-through rate, cost-per-click, position, and quality score) on bidding results over the three matching types, and showed that as the keyword matching types become narrower, almost all the keyword performance indices increase, but there is no significant change in cost per click. Du et al. (2017) constructed a hierarchical Bayesian framework to study the influence of keyword matching types on the advertising performance, and demonstrated the importance of differentiating bidding strategies over various matching types. Yang et al. (2021a) empirically explored the relationship between matching types and the advertising performance, and showed that exact match led to better performance than broad match.

**2.3 Budget Optimization**

How to efficiently allocate the limited advertising budget is the first and foremost problem faced by SSA advertisers (Yang et al., 2012; Yang and Xiong, 2020; Avadhanula et al., 2021; Yang et al., 2021b), which heavily affects other advertising decisions including keyword selection and keyword matching (Rusmevichientong and Williamson, 2006; Zhang et al., 2014; Yang et al., 2019).

With consideration of the entire life cycle of advertising campaigns, Yang et al. (2012) developed a hierarchical budget optimization framework (BOF) to handle budget decisions at three levels: allocation across search markets, temporal distribution over a series of slots (e.g., days) and adjustment of the remaining budget (e.g., daily budgeting), and presented a set of solution algorithms to efficiently solve identified budget decision problems in SSA. In a consequent work, Yang et al. (2014) proposed a dynamic multi-campaign budget planning strategy using optimal control techniques, with consideration of the substitution relationship between campaigns from three dimensions, i.e., campaign contents, promotional periods, and target regions, and showed that the overlapping degree between campaigns has serious effects on budgeting decisions and advertising performance.

With consideration of the budget constraint, the inventory constraint during the promotion season and the unknown relationship between the advertising expenditure and consequent sales, Yang and Xiong (2020) developed a nonparametric dynamic budget allocation strategy combining the exploration of learning the market sales response and the exploitation of planning the budget, and showed that their strategy was asymptotically optimal as the market size increases and achieved the near-best asymptotic performance in a worst-case. By explicitly considering the long-term influence of ad exposures to users, Hao et al. (2020) formulated a bi-level optimization framework for bidding strategy learning for each user and budget allocation among all users, and showed that their algorithm performed well in terms of cumulative revenue. Avadhanula et al. (2021) took the budget allocation across multiple platforms as a stochastic bandit with knapsacks problem, and developed an algorithm for both the discrete and continuous bid-spaces. Yang et al. (2021b) formulated the budget allocation as an optimal control problem with a generalized Vidale–Wolfe model (Yang et al., 2021c) as advertising dynamics, taking into account two useful indexes representing the advertising elasticity and the word-of-mouth (WoM) effect, and showed that their strategy outperforms other VW derivatives in terms of both payoff and ROI in either concave or S-shaped settings.

**2.4 Summary**

Table 2 summarizes the related literature on keyword selection and keyword matching from four aspects of techniques used, whether it considers keyword selection, includes the three matching types, and provides solutions.

**Table 2. Keyword Targeting Related Research**

| Reference | Techniques | Keyword Selection | Keyword Matching Types | Solution |
|---|---|---|---|---|
| Jones et al. (2006) | Linear regression, Linear SVMs, Decision trees | No | Broad match | Yes |
| Rusmevichientong and Williamson (2006) | Adaptive approximation | Yes | None | Yes |
| Kiritchenko and Jiline (2008) | Feature selection | Yes | None | Yes |
| Radlinski et al. (2008) | Support vector regression | No | Broad match and exact match | Yes |
| Singh and Roychowdhury (2008) | Graph model | No | Broad match | Yes |
| Even Dar et al. (2009) | Linear programming | No | Broad match and exact match | Yes |
| Mahdian and Wang (2009) | Approximation algorithm | No | Broad match | Yes |
| Gupta et al. (2009) | Max-margin voted perceptron | No | Broad match | Yes |
| Desai et al. (2014) | Game-theoretic model | Yes | None | No |
| Zhang et al. (2014) | Mixed integer optimization | Yes | None | Yes |
| Amaldoss et al. (2016) | Game-theoretic model | No | Broad match and exact match | No |
| Grbovic et al. (2016) | Distributed language model | No | Broad match | Yes |
| Du et al. (2017) | Hierarchical Bayesian model, MCMC | Yes | All three types | No |
| Lu and Yang (2017) | Hierarchical Bayesian model, MCMC | Yes | None | No |
| Ramaboa and Fish (2018) | Hypothesis testing | No | All three types | No |

| Arroyo-Cañada and Gil-Lafuente (2019) | Technique for order of preference by similarity to ideal solution (TOPSIS) | Yes | None | Yes |
|---|---|---|---|---|
| Yang et al. (2019) | Multilevel optimization | Yes | None | Yes |
| Nagpal and Petersen (2020) | Tobit model, Latent semantic analysis | Yes | None | Yes |
| Yang et al. (2021a) | Difference-in-difference-in-difference (DDD) | Yes | All three types | No |
| Symitsi et al. (2022) | Markowitz portfolio theory | Yes | None | Yes |
| This study | MCMC, Stochastic programming | Yes | All three types | Yes |

As illustrated in Table 2, prior research explored either keyword selection or keyword matching separately, while ignoring the necessity of addressing keyword selection and keyword matching problems in an integrated manner. This study fills the research gap by providing an effective optimization strategy for keyword targeting to select keywords and determine matching types simultaneously. To the best of our knowledge, this is the first study in this direction.

## 3. The Model and Solution

This section is split into two stages, i.e., the data distribution estimation for keyword performance indices and stochastic optimization for keyword targeting. In the first stage, we construct a data distribution estimation model and apply a Markov Chain Monte Carlo method (Chen et al., 2000) to make inference about unobserved keyword performance indices over the three matching types. In the second stage, based on the data estimation results, we construct a stochastic keyword targeting model combining operations of keyword selection and keyword matching, and develop a branch-and-bound algorithm incorporating a stochastic simulation process to solve our keyword targeting model. The notations used in this paper are listed in Table 3.

**Table 3. Notations**

| Notation | Definition |
|---|---|
| $d_{k,j,i}$ | The number of impressions of keyword $k$ in ad-group $j$ over matching type $i$ |
| $c_{k,j,i}$ | The click-through rate (CTR) of keyword $k$ in ad-group $j$ over matching type $i$ |
| $v_{k,j}$ | The value-per-click (VPC) of keyword $k$ in ad-group $j$ |
| $p_{k,j}$ | The cost-per-click (CPC) of keyword $k$ in ad-group $j$ |
| $\theta_j^d(\theta_j^c)$ | The 3-dimensional mean vector of a multivariate normal distribution for impressions (or CTRs) in ad-group $j$ |
| $\Sigma_j^d(\Sigma_j^c)$ | The $3 \times 3$ covariance matrix of a multivariate normal distribution for impressions (or CTRs) in ad-group $j$ |
| $\mu_0^d(\mu_0^c)$ | The prior mean of $\theta_j^d(\theta_j^c)$ for impressions (or CTRs) |
| $\Lambda_0^d(\Lambda_0^c)$ | The prior variance of $\theta_j^d(\theta_j^c)$ for impressions (or CTRs) |
| $\nu_0^d(\nu_0^c)$ | The prior scalar parameter of $\Sigma_j^d(\Sigma_j^c)$ for impressions (or CTRs) |
| $S_0^{d^{-1}}(S_0^{c^{-1}})$ | The prior matrix parameter of $\Sigma_j^d(\Sigma_j^c)$ for impressions (or CTRs) |
| $x_{k,j,i}$ | The binary decision variable indicating whether matching type $i$ is determined for keyword $k$ in ad-group $j$ or not |
| $B$ | The budget for a SSA campaign |
| $\alpha$ | The prescribed probability of the budget chance constraint |
| $n_j$ | The number of keywords in ad-group $j$ |
| $m$ | The number of ad-groups |

### 3.1 Data Distribution Estimation for Keyword Performance Indices

In SSA, advertisers choose a matching type for each selected keyword. In practice, keyword performance indices only over the chosen matching type can be observed in the promotional period. This results in a problem of incomplete information about performance indices over the keyword matching types. Thus, before making keyword targeting decisions, advertisers have to make inference about keyword performance indices over the three matching types. More specifically, given a set of keywords and performance indices for each keyword over a certain matching type in a promotional period, how to make inference about unobserved performance indices for each keyword over the other keyword matching types?

*3.1.1 Data Distribution Estimation Model*

We use a fully Bayesian model (Carrigan et al., 2007) to jointly simulate distributions of keyword performance indices with unobserved data over all the three matching types. Let $d_{k,j,i}$ denote the number of impressions of keyword $k$ ($k = 1,2,...,n$) in ad-group $j$ ($j = 1,2,...,m$) over matching type $i$ in a search market, $i = 1,2,3$ indicates exact match, phrase match and broad match, respectively. Let $c_{k,j,i}$ denote click-through rate (CTR) of keyword $k$ in ad-group $j$ over matching type $i$. Since cost-per-click (CPC) and value-per-click (VPC) are not significantly influenced by matching types (Ramaboa and Fish, 2018), we denote CPC and VPC of keyword $k$ in ad-group $j$ as $p_{k,j}$ and $v_{k,j}$, respectively.

Keywords in the same ad group focus on one or more common promotional products (or services)[3], and thus are closely related. Hence, we assume that performance indices (i.e., impression and CTR) for keywords in the same ad-group over the three matching types follow the same multivariate normal (MVN) distribution. Then the data distribution estimation problem for keyword performance indices can be transferred into the one for ad-group performance indices. Specifically, given an advertising campaign with $m$ ad-groups where there are $n_j$ keywords in ad-group $j$ ($j = 1, \dots, m$), we denote the number of impressions of keyword $k$ over the three matching types in ad-group $j$ as $d_{k,j} = (d_{k,j,1}, d_{k,j,2}, d_{k,j,3})^T$, $k = 1, \dots, n_j$.[4]

As impressions are non-negative ($d_{k,j,i} \geq 0$), we use the log function to transform impressions from the interval $[0, +\infty)$ into the real line $(-\infty, +\infty)$, i.e., $d'_{k,j} = (d'_{k,j,1}, d'_{k,j,2}, d'_{k,j,3})^T = \left(log(d_{k,j,1}), log(d_{k,j,2}), log(d_{k,j,3})\right)^T$.[5] We assume $d'_{k,j} \sim MVN(\theta_j^d, \Sigma_j^d)$, where $MVN(\theta_j^d, \Sigma_j^d)$ is a MVN distribution characterized by a 3-dimensional mean vector ($\theta_j^d$) and a $3 \times 3$ covariance matrix ($\Sigma_j^d$). Formally, the distribution estimation model for ad-group impressions is given as

$$d'_{k,j} \sim MVN(\theta_j^d, \Sigma_j^d),$$

$$\theta_j^d = \begin{bmatrix} E(d'_{j,1}) \\ E(d'_{j,2}) \\ E(d'_{j,3}) \end{bmatrix},$$

$$\Sigma_j^d = \begin{bmatrix} \Sigma_{j,11}^d & \Sigma_{j,12}^d & \Sigma_{j,13}^d \\ \Sigma_{j,21}^d & \Sigma_{j,22}^d & \Sigma_{j,23}^d \\ \Sigma_{j,31}^d & \Sigma_{j,32}^d & \Sigma_{j,33}^d \end{bmatrix},$$

$$k = 1, 2, \dots, n_j, \ j = 1, 2, \dots, m. \qquad (1)$$

---

[3] https://support.google.com/google-ads/answer/2375404?hl=en
[4] In the rest of this paper we use "the number of impressions" and "impressions" interchangeably.
[5] Since the log function is undefined at zero, when the number of impressions is equal to zero, we add a very small positive number close to zero to it to ensure the smooth operation in the transformation.

In Bayesian statistics, the conjugate prior of the mean vector $\theta_j^d$ of a MVN is another MVN, while the conjugate prior of the covariance matrix $\Sigma_j^d$ of a MVN is an inverse-Wishart distribution, which are given as

$$\theta_j^d \sim MVN(\mu_0^d, \Lambda_0^d), \quad (2)$$

$$\Sigma_j^d \sim inverseWishart(v_0^d, S_0^{d^{-1}}), \quad (3)$$

where $\mu_0^d$ and $\Lambda_0^d$ are the prior mean and variance of $\theta_j^d$ for impressions, respectively; $v_0^d$ is the degree of freedom and $S_0^{d^{-1}}$ is the scale matrix.

Analogously, we denote the CTR for keyword $k$ over the three matching types $i = 1,2,3$ in ad-group $j$ ($j = 1, \ldots, m$) as $c_{k,j} = (c_{k,j,1}, c_{k,j,2}, c_{k,j,3})^T$, $k = 1, \ldots, n_j$. We use the logit function (i.e., the inverse of the sigmoid function) to transform the CTR values from the interval $[0,1]$ into the real line $(-\infty, +\infty)$.[6] Then $c'_{k,j} = (c'_{k,j,1}, c'_{k,j,2}, c'_{k,j,3})^T = \left(logit(c_{k,j,1}), logit(c_{k,j,2}), logit(c_{k,j,3})\right)^T$ obeys a MVN distribution $MVN(\theta_j^c, \Sigma_j^c)$, where $\theta_j^c = (E[c'_{j,1}], E[c'_{j,2}], E[c'_{j,3}])^T$ is a 3-dimensional mean vector, and $\Sigma_j^c$ is a $3 \times 3$ covariance matrix for CTRs. Formally, the distribution estimation model for ad-group CTRs is given as

$$c'_{k,j} \sim MVN(\theta_j^c, \Sigma_j^c),$$

$$\theta_j^c = \begin{bmatrix} E(c'_{j,1}) \\ E(c'_{j,2}) \\ E(c'_{j,3}) \end{bmatrix},$$

$$\Sigma_j^c = \begin{bmatrix} \Sigma_{j,11}^c & \Sigma_{j,12}^c & \Sigma_{j,13}^c \\ \Sigma_{j,21}^c & \Sigma_{j,22}^c & \Sigma_{j,23}^c \\ \Sigma_{j,31}^c & \Sigma_{j,32}^c & \Sigma_{j,33}^c \end{bmatrix},$$

$$k = 1,2, \ldots, n_j, \ j = 1,2, \ldots, m. \quad (4)$$

$$\theta_j^c \sim MVN(\mu_0^c, \Lambda_0^c). \quad (5)$$

$$\Sigma_j^c \sim inverseWishart(v_0^c, S_0^{c^{-1}}). \quad (6)$$

---

[6] Since the logit function is undefined at zero and one, when the CTR is equal to zero, we add a very small positive number close to zero to it; when the CTR is equal to one, we subtract a very small positive number close to zero from it.

Based on the above distribution estimation models for impressions and CTRs, we apply the Markov Chain Monte Carlo method (MCMC) method (Gamerman and Lopes, 2006) to make inference about the model parameters and unobserved performance indices.

*3.1.2 Gibbs Sampling*

The process of solving the distribution estimation models for impressions and CTRs through Gibbs sampling are similar. In the following, we take impressions to illustrate the process. To simplify notations, we use $d_j = \{d_{1,j}, d_{2,j}, \ldots, d_{n_j,j}\}$ to represent the log-transformed impressions[7]. For ad-group $j$, the distribution of the mean vector $\theta_j^d$ is

$$p(\theta_j^d) = MVN(\mu_0^d, \Lambda_0^d)$$
$$= (2\pi)^{-3/2} |\Lambda_0^d|^{-1/2} \exp\{-\frac{1}{2}(\theta_j^d - \mu_0^d)^T \Lambda_0^{d-1}(\theta_j^d - \mu_0^d)\}$$
$$= (2\pi)^{-3/2} |\Lambda_0^d|^{-1/2} \exp\{-\frac{1}{2}\theta_j^{d^T} \Lambda_0^{d-1} \theta_j^d + \theta_j^{d^T} \Lambda_0^{d-1} \mu_0^d - \frac{1}{2}\mu_0^{d^T} \Lambda_0^{d-1} \mu_0^d\}$$
$$\propto \exp\{-\frac{1}{2}\theta_j^{d^T} \Lambda_0^{d-1} \theta_j^d + \theta_j^{d^T} \Lambda_0^{d-1} \mu_0^d\}$$
$$= \exp\{-\frac{1}{2}\theta_j^{d^T} A_0^d \theta_j^d + \theta_j^{d^T} b_0^d\}, \qquad (7)$$

where $A_0^d = \Lambda_0^{d-1}$ and $b_0^d = \Lambda_0^{d-1} \mu_0^d$.

Equation (7) implies that, if $\theta_j^d$ has a density on $\mathbb{R}^3$ that is proportional to $exp\{-\frac{1}{2}\theta_j^{d^T} A_0^d \theta_j^d + \theta_j^{d^T} b_0^d\}$, then it must have a MVN with covariance $A_0^{d-1}$ and mean $A_0^{d-1} b_0^d$.

In our sampling model, keyword impressions in ad-group $j$, i.e., $\{d_{1,j}, \ldots, d_{n_j,j}\}$, are independent and identically distributed $MVN(\theta_j^d, \Sigma_j^d)$. Then the joint sampling density of impressions is given as

$$p\left(d_{1,j}, \ldots, d_{n_j,j} \middle| \theta_j^d, \Sigma_j^d\right)$$
$$= \prod_{k=1}^{n_j} (2\pi)^{-3/2} |\Sigma_j^d|^{-1/2} \exp\{-\frac{1}{2}(d_{k,j} - \theta_j^d)^T \Sigma_j^{d-1}(d_{k,j} - \theta_j^d)\}$$
$$= (2\pi)^{-3n_j/2} |\Sigma_j^d|^{-n_j/2} \exp\{-\frac{1}{2}\sum_{k=1}^{n_j}(d_{k,j} - \theta_j^d)^T \Sigma_j^{d-1}(d_{k,j} - \theta_j^d)\}$$
$$\propto \exp\{-\frac{1}{2}\theta_j^{d^T} A_1^d \theta_j^d + \theta_j^{d^T} b_1^d\}, \qquad (8)$$

---

[7] In the model and experiments, estimated impressions and CTRs are transformed back from $(-\infty, +\infty)$ to $[0, +\infty)$ and $[0,1]$ using the inverse functions, separately.

where $A_1^d = n\Sigma_j^{d^{-1}}$, $b_1^d = n\Sigma_j^{d^{-1}}\bar{d}_j$, and $\bar{d}_j$ is the vector of impression-specific averages over the three matching types, i.e.,

$$\bar{d}_j = (\frac{1}{n_j}\sum_{k=1}^{n_j} d_{k,j,1}, \frac{1}{n_j}\sum_{k=1}^{n_j} d_{k,j,2}, \frac{1}{n_j}\sum_{k=1}^{n_j} d_{k,j,3})^T. \quad (9)$$

Combining Equations (7) and (8) yields

$$p\left(\theta_j^d \big| d_{1,j}, \ldots, d_{n_j,j}, \Sigma_j^d\right)$$

$$\propto \exp\{-\frac{1}{2}\theta_j^{d^T} A_0^d \theta_j^d + \theta_j^{d^T} b_0^d\} \times \exp\{-\frac{1}{2}\theta_j^{d^T} A_1^d \theta_j^d + \theta_j^{d^T} b_1^d\}$$

$$= \exp\{-\frac{1}{2}\theta_j^{d^T} A_n^d \theta_j^d + \theta_j^{d^T} b_n^d\}, \quad (10)$$

where $A_n^d = A_0^d + A_1^d = \Lambda_0^{d^{-1}} + n\Sigma_j^{d^{-1}}$ and $b_n^d = b_0^d + b_1^d = \Lambda_0^{d^{-1}}\mu_0^d + n\Sigma_j^{d^{-1}}\bar{d}_j$.

Equation (10) implies that the conditional distribution of the mean vector for impressions ($\theta_j^d$) must be a MVN with covariance $A_n^{d^{-1}}$ and mean $A_n^{d^{-1}} b_n^d$. Then we have

$$E\left[\theta_j^d \big| d_{1,j}, \ldots, d_{n_j,j}, \Sigma_j^d\right] = \mu_n^d = \left(\Lambda_0^{d^{-1}} + n\Sigma_j^{d^{-1}}\right)^{-1}\left(\Lambda_0^{d^{-1}}\mu_0^d + n\Sigma_j^{d^{-1}}\bar{d}_j\right), \quad (11)$$

$$Cov\left[\theta_j^d \big| d_{1,j}, \ldots, d_{n_j,j}, \Sigma_j^d\right] = \Lambda_n^d = \left(\Lambda_0^{d^{-1}} + n\Sigma_j^{d^{-1}}\right)^{-1}, \quad (12)$$

$$p\left(\theta_j^d \big| d_{1,j}, \ldots, d_{n_j,j}, \Sigma_j^d\right) = MVN(\mu_n^d, \Lambda_n^d), \quad (13)$$

The density of covariance matrix $\Sigma_j^d$ is given by the inverse-Wishart ($v_0^d, S_0^{d^{-1}}$), i.e,

$$p(\Sigma_j^d) = \left[2^{\frac{3v_0^d}{2}} \pi^{\frac{3}{2}} |S_0^d|^{-\frac{v_0^d}{2}} \prod_{i=1}^{3} \Gamma\left(\frac{[v_0^d+1-i]}{2}\right)\right]^{-1} \times |\Sigma_j^d|^{-(v_0^d+4)/2} \times \exp\{-\frac{1}{2}tr(S_0^d \Sigma_j^{d^{-1}})\}. \quad (14)$$

To combine the above prior distribution $p(\Sigma_j^d)$ with the sampling distribution for impressions, we can rewrite the sampling distribution in Equation (8) as

$$p\left(d_{1,j}, \ldots, d_{n_j,j} \big| \theta_j^d, \Sigma_j^d\right)$$

$$= (2\pi)^{-\frac{3n_j}{2}} |\Sigma_j^d|^{-\frac{n_j}{2}} \exp\left\{-\frac{1}{2}\sum_{k=1}^{n_j}(d_{k,j} - \theta_j^d)^T \Sigma_j^{d^{-1}}(d_{k,j} - \theta_j^d)\right\}$$

$$= (2\pi)^{-\frac{3n_j}{2}} |\Sigma_j^d|^{-\frac{n_j}{2}} \exp\left\{-\frac{1}{2}tr(S_\theta^d \Sigma_j^{d^{-1}})\right\}, \quad (15)$$

where $S_\theta^d = \sum_{k=1}^{n_j}(d_{k,j} - \theta_j^d)(d_{k,j} - \theta_j^d)^T$.

Combining Equations (14) and (15), we have the conditional distribution of the covariance matrix $\Sigma_j^d$, given as

$$p\left(\Sigma_j^d \middle| d_{1,j}, \ldots, d_{n_j,j}, \theta_j^d\right)$$

$$\propto p(\Sigma_j^d) \times p\left(d_{1,j}, \ldots, d_{n_j,j} \middle| \theta_j^d, \Sigma_j^d\right)$$

$$\propto \left(|\Sigma_j^d|^{-(v_0^d+4)/2} \times \exp\{-\tfrac{1}{2}tr(S_0^d \Sigma_j^{d^{-1}})\}\right) \times \left(|\Sigma_j^d|^{-\tfrac{n_j}{2}} \exp\{-\tfrac{1}{2}tr(S_\theta^d \Sigma_j^{d^{-1}})\}\right)$$

$$= |\Sigma_j^d|^{-(v_0^d+n_j+4)/2} \exp\{-\tfrac{1}{2}tr([S_0^d + S_\theta^d]\Sigma_j^{d^{-1}})\}. \qquad (16)$$

Thus, we have

$$\{\Sigma_j^d | d_{1,j}, \ldots, d_{n_j,j}, \theta_j^d\} \sim inverseWishart(v_0^d + n_j, [S_0^d + S_\theta^d]^{-1}). \qquad (17)$$

Let $O_{k,j} = (O_{k,j,1}, O_{k,j,2}, O_{k,j,3})^T$ be a binary vector where $O_{k,j,i} = 1$ implies that in ad-group $j$, the number of impressions of keyword $k$ over matching type $i$ (i.e., $d_{k,j,i}$) is observed, whereas $O_{k,j,i} = 0$ implies that $d_{k,j,i}$ is unobserved. In SSA, we can assume that unobserved impressions are missing at random, i.e., $O_{k,j}$ and $d_{k,j}$ are statistically independent, and that the distribution of $O_{k,j}$ does not depend on $\theta_j^d$ or $\Sigma_j^d$. The sampling probability for impressions from keyword $k$ in ad-group $j$ is $p(O_{k,j})$ multiplied by the marginal probability of observed impressions after integrating out unobserved impressions, i.e.,

$$p(O_{k,j}, \{d_{k,j,i}: O_{k,j,i} = 1\} | \theta_j^d, \Sigma_j^d)$$

$$= p(O_{k,j}) \times p(\{d_{k,j,i}: O_{k,j,i} = 1\} | \theta_j^d, \Sigma_j^d)$$

$$= p(O_{k,j}) \times \int \left\{ p(d_{k,j,1}, d_{k,j,2}, d_{k,j,3}) | \theta_j^d, \Sigma_j^d) \prod_{d_{k,j,i}: O_{k,j,i}=0} d(d_{k,j,i}) \right\}. \qquad (18)$$

The integration in Equation (18) can be done using Gibbs sampling to make inference about $\theta_j^d$, $\Sigma_j^d$ and impressions over unobserved keyword matching types.

Let $d_j$ be a $n_j \times 3$ matrix of the observed and unobserved impressions over the three matching types in ad-group $j$, $O_j$ be a $n_j \times 3$ matrix. Then $d_j$ can be assumed to be composed of two parts: one is the observed impressions $d_j^{obs} = \{d_{k,j,i}: O_{k,j,i} = 1\}$; another is the unobserved impressions $d_j^{unobs} = \{d_{k,j,i}: O_{k,j,i} = 0\}$. To approximate the posterior distribution of unknown and unobserved quantities $p(\theta_j^d, \Sigma_j^d, d_j^{unobs} | d_j^{obs})$, we build a Gibbs sampling scheme as follows. Given starting values $\{\Sigma_j^{d(0)}, d_j^{unobs(0)}\}$, we generate $\{\theta_j^{d(s+1)}, \Sigma_j^{d(s+1)}, d_j^{unobs(s+1)}\}$ from $\{\theta_j^{d(s)}, \Sigma_j^{d(s)}, d_j^{unobs(s)}\}$ through the following process:

1) sampling $\theta_j^{d(s+1)}$ from $p\left(\theta_j^d | d_j^{obs}, d_j^{unobs(s)}, \Sigma_j^{d(s)}\right)$;

2) sampling $\Sigma_j^{d\,(s+1)}$ from $p\left(\Sigma_j^d | d_j^{obs}, d_j^{unobs\,(s)}, \theta_j^{d\,(s+1)}\right)$;

3) sampling $d_j^{unobs\,(s+1)}$ from $p\left(d_j^{unobs} | d_j^{obs}, \theta_j^{d\,(s+1)}, \Sigma_j^{d\,(s+1)}\right)$.

In steps 1 and 2, for ad-group $j$, the fixed observed impressions ($d_j^{obs}$) is combined with the current sampled unobserved impressions ($d_j^{unobs\,(s)}$) to shape a complete matrix of impressions in the $s$-th round of sampling ($d_j^{(s)}$). The $n_j$ rows of the impression matrix ($d_j^{(s)}$) can be plugged into Equations (13) and (17) to obtain the full conditional distributions of the MVN parameters ($\theta_j^d$ and $\Sigma_j^d$). In Step 3, we sample the unobserved elements conditional on the observed elements of the keyword impression matrix:

$$p(d_j^{unobs} | d_j^{obs}, \theta_j^d, \Sigma_j^d)$$
$$\propto p(d_j^{unobs}, d_j^{obs} | \theta_j^d, \Sigma_j^d)$$
$$= \prod_{k=1}^{n_j} p(d_{k,j}^{unobs}, d_{k,j}^{obs} | \theta_j^d, \Sigma_j^d)$$
$$\propto \prod_{k=1}^{n_j} p(d_{k,j}^{unobs} | d_{k,j}^{obs}, \theta_j^d, \Sigma_j^d). \qquad (19)$$

This is made possible through the following result about MVNs. Let $d_j \sim MVN(\theta_j^d, \Sigma_j^d)$, $a$ be a subset of indices $\{1,2,3\}$, and $b$ be the complement of $a$. According to inverses of partitioned matrices, we have

$$\{d_{k,j_{[b]}} | d_{k,j_{[a]}}, \theta_j^d, \Sigma_j^d\} \sim MVN(\theta_{j\,b|a}^d, \Sigma_{j\,b|a}^d),$$
$$\theta_{j\,b|a}^d = \theta_{j\,[b]}^d + \Sigma_{j\,[b,a]}^d \left(\Sigma_{j\,[a,a]}^d\right)^{-1} (d_{k,j_{[a]}} - \theta_{j\,[a]}^d),$$
$$\Sigma_{j\,b|a}^d = \Sigma_{j\,[b,b]}^d - \Sigma_{j\,[b,a]}^d \left(\Sigma_{j\,[a,a]}^d\right)^{-1} \Sigma_{j\,[a,b]}^d,$$
$$k = 1,2,\ldots,n_j, \qquad (20)$$

where $d_{k,j_{[b]}}$ represents the elements of $d_{k,j}$ corresponding to indices in $b$, $\theta_{j\,[b]}^d$ represents the elements of $\theta_j^d$ corresponding to indices in $b$, and $\Sigma_{j\,[a,b]}^d$ is the matrix made up of elements that are in rows $a$ and columns $b$ of $\Sigma_j^d$.

## 3.2 Stochastic Keyword Targeting Optimization

Based on the data estimation results of keyword performance indices over the three matching types discussed in Section 3.1, we construct a stochastic model for keyword targeting to maximize the expected profit from SSA campaigns. The decision scenario of keyword targeting under consideration by this research is described as follows: given a set of keywords, how to select appropriate keywords and determine matching types for these selected keywords simultaneously?

*3.2.1 The Objective of Keyword Targeting Optimization*

The decision variable $x_{k,j,i}$ indicates whether matching type $i$ is chosen for keyword $k$ in ad-group $j$ or not. Let $x_{k,j,0} = 1$ denote the case when keyword $k$ in ad-group $j$ is not chosen over any of the three matching types. Then we have $\sum_{i=0}^{3} x_{k,j,i} = 1$. As a matter of fact, $i = 0$ means that keyword $k$ is not selected by the advertiser, i.e., $d_{k,j,0} = 0$ and $c_{k,j,0} = 0$. For a SSA campaign, the cost is $\sum_{j=1}^{m} \sum_{k=1}^{n} \sum_{i=0}^{3} x_{k,j,i} d_{k,j,i} c_{k,j,i} p_{k,j}$, and the expected profit is

$$z(x_{k,j,i}) = \sum_{j=1}^{m} \sum_{k=1}^{n} \sum_{i=0}^{3} x_{k,j,i} d_{k,j,i} c_{k,j,i} (v_{k,j} - p_{k,j}). \qquad (21)$$

In this research, we use the data distribution estimation of performance indices for an ad-group to approximate those for keywords in that ad-group. We regard $d_{k,j,i}$ and $c_{k,j,i}$ as random vectors capturing the uncertainty in SSA. Then $z(x_{k,j,i})$ is also a random variable. The keyword targeting decision aims to maximize the expected profit generated from a SSA campaign, which is given as

$$max\ E[z(x_{k,j,i})] = E\left[\sum_{j=1}^{m} \sum_{k=1}^{n} \sum_{i=0}^{3} x_{k,j,i} d_{k,j,i} c_{k,j,i} (v_{k,j} - p_{k,j})\right]. \qquad (22)$$

*3.2.2 The Chance Constraint of the Budget*

It is naturally assumed that the available budget for an advertiser is relatively less than a sufficient amount. In a SSA campaign, let $B > 0$ denote the advertising budget for a SSA campaign, then we have

$$\sum_{j=1}^{m} \sum_{k=1}^{n} \sum_{i=0}^{3} x_{k,j,i} d_{k,j,i} c_{k,j,i} p_{k,j} \leq B. \qquad (23)$$

Due to the stochastic nature of $d_{k,j,i}$ and $c_{k,j,i}$, we can use the chance constraint of the budget to control the cost of a SSA campaign. Specifically, the probability that the cost of a SSA campaign is less than the available budget is larger than or equal to a specific confidence level $\alpha$, i.e.,

$$P\left\{\sum_{j=1}^{m} \sum_{k=1}^{n} \sum_{i=0}^{3} x_{k,j,i} d_{k,j,i} c_{k,j,i} p_{k,j} \leq B\right\} \geq \alpha. \qquad (24)$$

*3.2.3 The Stochastic Keyword Targeting Model*

In summary, based on the data distribution estimation for keyword performance indices discussed in Section 3.1, we formulate the keyword targeting problem as a stochastic optimization model, given as

$$\max E\left[\sum_{j=1}^{m}\sum_{k=1}^{n}\sum_{i=0}^{3} x_{k,j,i} d_{k,j,i} c_{k,j,i} (v_{k,j} - p_{k,j})\right],$$

$$\text{s.t. } P\left\{\sum_{j=1}^{m}\sum_{k=1}^{n}\sum_{i=0}^{3} x_{k,j,i} d_{k,j,i} c_{k,j,i} p_{k,j} \leq B\right\} \geq \alpha,$$

$$\sum_{i=0}^{3} x_{k,j,i} = 1,$$

$$\left(\log(d_{k,j,1}), \log(d_{k,j,2}), \log(d_{k,j,3})\right)^T \sim MVN\left(\theta_j^d, \Sigma_j^d\right),$$

$$\left(logit(c_{k,j,1}), logit(c_{k,j,2}), logit(c_{k,j,3})\right)^T \sim MVN\left(\theta_j^c, \Sigma_j^c\right),$$

$$\theta_j^d = \begin{bmatrix} E(d'_{j,1}) \\ E(d'_{j,2}) \\ E(d'_{j,3}) \end{bmatrix},$$

$$\Sigma_j^d = \begin{bmatrix} \Sigma_{j,11}^d & \Sigma_{j,12}^d & \Sigma_{j,13}^d \\ \Sigma_{j,21}^d & \Sigma_{j,22}^d & \Sigma_{j,23}^d \\ \Sigma_{j,31}^d & \Sigma_{j,32}^d & \Sigma_{j,33}^d \end{bmatrix},$$

$$\theta_j^c = \begin{bmatrix} E(c'_{j,1}) \\ E(c'_{j,2}) \\ E(c'_{j,3}) \end{bmatrix},$$

$$\Sigma_j^c = \begin{bmatrix} \Sigma_{j,11}^c & \Sigma_{j,12}^c & \Sigma_{j,13}^c \\ \Sigma_{j,21}^c & \Sigma_{j,22}^c & \Sigma_{j,23}^c \\ \Sigma_{j,31}^c & \Sigma_{j,32}^c & \Sigma_{j,33}^c \end{bmatrix},$$

$$x_{k,j,i} = 0/1, d_{k,j,0} = 0, c_{k,j,0} = 0, v_{k,j} \geq 0, p_{k,j} \geq 0,$$

$$k = 1,2,\ldots,n_j, i = 0,1,2,3, j = 1,2,\ldots,m. \quad (25)$$

In our stochastic keyword targeting model (Equation 25), the decision variable $x_{k,j,i}$ is 0-1 binary. The objective function is to maximize the expected profit from a SSA campaign. The first constraint is the chance constraint of advertiser's budget. The second constraint limits each keyword being selected over only one matching type or not selected at all. In addition, we also state the estimated data distributions and non-negative performance indices.

*3.2.4 BB-KSM Algorithm*

The stochastic keyword targeting model can be solved by branch-and-bound algorithm combined with stochastic simulation. The branch-and-bound consists of a systematic enumeration of feasible solutions which helps reduce computational efforts and find the optimal keyword targeting result (Kosuch and Lisser, 2010). In particular, given a set of keywords and a fixed budget, based on the estimated distribution of keyword performance indices over the three matching types, an optimal solution should adaptively select appropriate keywords and determine matching types to maximize the expected profit in an uncertain market. Since a same keyword in different ad-groups can have different matching types, which leads to different performance indices. Thus, we regard an identical keyword in different ad-groups as different keywords. As each keyword can only choose no more than one matching type, we treat the keyword-matching combinations as the basic unit and develop a branch-and-bound algorithm called BB-KSM (i.e., the branch-and-bound for keyword selection and matching) to solve our stochastic keyword targeting model. In the following, we first describe the stochastic simulation process in the branch-and-bound. Then we explain how to calculate the upper bound for BB-KSM. Finally, we present the BB-KSM algorithm.

First, the stochastic simulation process is used to confirm whether the chance constraint is satisfied in the branch-and-bound process, i.e., the probability that the current cost for selected keywords has not exceeded campaign budget is above the acceptable threshold value. Specifically, in ad-group $j$, once keyword $k$ is selected over matching type $i$, we sample random variables from their estimated distributions and check whether the chance constraint of the budget $P\{\sum_{j=1}^{m}\sum_{k=1}^{n}\sum_{i=0}^{3} x_{k,j,i} d_{k,j,i} c_{k,j,i} p_{k,j} \leq B\} \geq \alpha$ is satisfied, which is performed in lines from 4 to 8 of the pseudocode of the BB-KSM algorithm.

Next, to obtain the upper bound (i.e., SUP) for BB-KSM, we relax $x_{k,j,i}$ from a 0-1 binary variable to a continuous variable in the interval of [0,1] in the stochastic keyword targeting model (Equation 25). Under the continuous relaxation, the chance constraint of the budget $P\{\sum_{j=1}^{m}\sum_{k=1}^{n}\sum_{i=0}^{3} x_{k,j,i} d_{k,j,i} c_{k,j,i} p_{k,j} \leq B\} \geq \alpha$ defines a convex set if $\sum_{j=1}^{m}\sum_{k=1}^{n}\sum_{i=0}^{3} x_{k,j,i} d_{k,j,i} c_{k,j,i} p_{k,j}$ is quasi-convex and $s_{k,j,i} = d_{k,j,i} c_{k,j,i} p_{k,j}$ has a log-concave density (Prekopa, 1995). The first condition is well satisfied since the left-hand side of the inequation $\sum_{j=1}^{m}\sum_{k=1}^{n}\sum_{i=0}^{3} x_{k,j,i} d_{k,j,i} c_{k,j,i} p_{k,j}$ is quasi-convex due to its linear formulation. As for the second condition, according to Cholette et al. (2012), the number of clicks from a keyword is binomial distributed with parameters $(d_{k,j,i}, c_{k,j,i})$, which can be accurately approximated by the normal distribution provided that $d_{k,j,i} \cdot c_{k,j,i} \geq 10$ and $d_{k,j,i} \cdot (1 - c_{k,j,i}) \geq 10$. In general, we could assume that $d_{k,j,i}$ and $c_{k,j,i}$ reasonably well meet these two inequalities. For a keyword, the cost is equal to the product of the number of clicks (the random variable) and the average cost per click (constant), which also obeys the normal distribution. Thus, the second condition can be satisfied. Consequently, after the continuous relaxation, the chance constraint of the budget $P\{\sum_{j=1}^{m}\sum_{k=1}^{n}\sum_{i=0}^{3}x_{k,j,i}d_{k,j,i}c_{k,j,i}p_{k,j} \leq B\} \geq \alpha$ defines a convex set with normally distributed keyword costs.

Then the continuous relaxed stochastic keyword targeting model (25) can be reformulated as an equivalent, deterministic second-order-cone-programming (SOCP) problem (Lobo et al., 1998). Through sampling from estimations for impressions and CTR, we can get the mean $\mu^{(s)}$ and variance $\sigma^{(s)^2}_{k,j,i}$ of the cost $s_{k,j,i}$ for each keyword. As the budget $B$ is a fixed constant, the inequality constraint $\sum_{j=1}^{m}\sum_{k=1}^{n}\sum_{i=0}^{3}x_{k,j,i}s_{k,j,i} \leq B$ is equal to

$$\frac{\sum_{j=1}^{m}\sum_{k=1}^{n}\sum_{i=0}^{3}x_{k,j,i}s_{k,j,i}-B-\left(\sum_{j=1}^{m}\sum_{k=1}^{n}\sum_{i=0}^{3}x_{k,j,i}E[s_{k,j,i}]-B\right)}{\sqrt{\sum_{j=1}^{m}\sum_{k=1}^{n}\sum_{i=0}^{3}x_{k,j,i}^2 Var[s_{k,j,i}]}} \leq -\frac{\sum_{j=1}^{m}\sum_{k=1}^{n}\sum_{i=0}^{3}x_{k,j,i}E[s_{k,j,i}]-B}{\sqrt{\sum_{j=1}^{m}\sum_{k=1}^{n}\sum_{i=0}^{3}x_{k,j,i}^2 Var[s_{k,j,i}]}}, \quad (26)$$

where the left side of the inequality represents a standard normal variant.

Thus, the chance constraint of the budget $P\{\sum_{j=1}^{m}\sum_{k=1}^{n}\sum_{i=0}^{3}x_{k,j,i}d_{k,j,i}c_{k,j,i}p_{k,j} \leq B\} \geq \alpha$ becomes

$$P\left\{\eta \leq -\frac{\sum_{j=1}^{m}\sum_{k=1}^{n}\sum_{i=0}^{3}x_{k,j,i}E[s_{k,j,i}]-B}{\sqrt{\sum_{j=1}^{m}\sum_{k=1}^{n}\sum_{i=0}^{3}x_{k,j,i}^2 Var[s_{k,j,i}]}}\right\} \geq \alpha, \quad (27)$$

where $\eta$ follows a standard normal distribution.

Then, the chance constraint of the budget can be reformulated as

$$\phi^{-1}(\alpha) \leq -\frac{\sum_{j=1}^{m}\sum_{k=1}^{n}\sum_{i=0}^{3}x_{k,j,i}E[s_{k,j,i}]-B}{\sqrt{\sum_{j=1}^{m}\sum_{k=1}^{n}\sum_{i=0}^{3}x_{k,j,i}^2 Var[s_{k,j,i}]}}$$

$$\Rightarrow \sum_{j=1}^{m}\sum_{k=1}^{n}\sum_{i=0}^{3}x_{k,j,i}E[s_{k,j,i}] + \phi^{-1}(\alpha)\sqrt{\sum_{j=1}^{m}\sum_{k=1}^{n}\sum_{i=0}^{3}x_{k,j,i}^2 Var[s_{k,j,i}]} \leq B,$$

$$\Rightarrow \sum_{j=1}^{m}\sum_{k=1}^{n}\sum_{i=0}^{3}x_{k,j,i}\mu^{(s)}_{k,j,i} + \phi^{-1}(\alpha)\sqrt{\sum_{j=1}^{m}\sum_{k=1}^{n}\sum_{i=0}^{3}x_{k,j,i}^2 \sigma^{(s)^2}_{k,j,i}} \leq B. \quad (28)$$

Consequently, we obtain a convex optimization model which can be solved by the interior point method (Wächter and Biegler, 2006). Its solution can be used as upper bound (SUP) in the branch-and-bound process.

In the following we develop a branch-and-bound algorithm (i.e., BB-KSM) to solve our stochastic keyword targeting model. The BB-KSM algorithm systematically enumerates the candidate solutions for optimal keyword targeting, where a set of candidate solutions is thought of as forming a rooted tree with the full set at the root. It explores branches of the tree, i.e., subsets of keyword targeting solutions. Before enumerating the candidate solutions of a branch, the branch is checked against upper (SUP) and lower (INF) estimated bounds on the optimum, and is discarded if it cannot produce a better keyword targeting solution than the existing best one found so far by the BB-KSM algorithm. In the exploring process, we sort keywords by the decreasing profit, then select keywords one-by-one and use stochastic simulation to check whether the chance constraint of the budget is still satisfied when selecting the next keyword. The overall framework and details of BB-KSM algorithm are given as follows.

| The BB-KSM Algorithm |
|---|
| **Input:** |
| $\{i\|i = 1,2,3\}$ – The three keyword matching types (i.e., exact match, phrase match and broad match) |
| $\{j\|j = 1,2,\dots,m\}$ –Ad-group set |
| $\{k\|k = 1,2,\dots,n_j\}$ – The keyword set in ad-group $j$ |
| $B$ – The budget constraint for the SSA campaign |
| $\alpha$– The acceptable probability for the chance constraint of the budget |
| $\theta_j^d, \Sigma_j^d$ – The mean vector and covariance matrix for keyword impressions in ad-group $j$ over the three matching types |
| $\theta_j^c, \Sigma_j^c$ – The mean vector and covariance matrix for the keyword CTRs in ad-group $j$ over the three matching types |
| $v_{k,j}$ – The VPC of keyword $k$ in ad-group $j$ |
| $p_{k,j}$ – The CPC of keyword $k$ in ad-group $j$ |
| **Output:** |
| $x_{k,j,i}$ – The binary decision variable for keyword targeting |
| **Procedure:** |
| Sort keywords in the order decreasing of profit, $x_{k,j,i} = 0$, Keyword_Targeting_List = $\emptyset$. |
| 1: **For** ad-group $j$ from 1 to $m$ |
| 2:    **For** keyword $k$ from 1 to $n_j$ |

> 3:     **For** matching type $i$ from 1 to 3
> 4:         $t' = 0$ and $x_{k,j,i} = 1$
> 5:         Extract $d_{k,j,i}$ and $c_{k,j,i}$ samples from $MVN(\theta_j^d, \Sigma_j^d)$ and $MVN(\theta_j^c, \Sigma_j^c)$
> 6:         **If** $\sum_{j=1}^{m}\sum_{k=1}^{n}\sum_{i=0}^{3} x_{k,j,i} d_{k,j,i} c_{k,j,i} p_{k,j} \leq B$ **then** $t'++$
> 7:         **Repeat** steps 5 and 6 for $t$ times, and $\alpha' = t'/t$
> 8:         **If** $\alpha' \geq \alpha$ and $\sum_{i=1}^{3} x_{k,j,i} \leq 1$, **then** INF = max{the expected profit}, add the feasible solution to Keyword_Targeting_List, and SUP $= \infty$; else $x_{k,j,i} = 0$
> 9:     End for
> 10:    End for
> 11: End for
> 12: **If** Keyword_Targeting_List $= \emptyset$, **then** go to step 16; **else** current_solution = solution in Keyword_Targeting_List with max{the expected profit}, go to step 13.
> 13: **If** SUP > INF for current_solution, **then** go to step 14; **else** delete the solution from Keyword_Targeting_List and go to step 12.
> 14: **If** there is no accepted keyword-matching left in the targeting solution that does not already have a plunged or rejected subset, **then** delete the solution from Keyword_Targeting_List, go to step 12; **else** following the ranking, choose the first accepted keyword-matching that does not already have a plunged or rejected subset and calculate SUP for the subset defined by rejecting this keyword-matching, and go to step 15.
> 15: **If** SUP ≤ INF, **then** delete this subset, go to step 14; **else** plunge the subtree as described in steps 1-11 and add the found branch together with SUP to Keyword_Targeting_List, **if** the expected profit of this solution > INF, **then** update INF, go to step 12.
> 16: **Return** the result of keyword targeting $x$.

## 4. Experimental Validation

### 4.1 Data Descriptions

We conduct a series of computational experiments to validate our keyword targeting strategy (i.e., BB-KSM) using a realworld dataset. The purpose of our experiments is two-fold. First, we aim to evaluate the performance of our keyword targeting strategy by comparing with seven baselines, in terms of the expected profit and the number of selected keywords. Second, we are intended to evaluate the effectiveness of our distribution estimation method described in Section 3.1.

Our experimental dataset is collected from field reports and SSA logs of advertising campaigns performance of an e-commerce retailer selling gift cards and Christmas presents. The dataset records the daily performance at the keyword level from September 2011 to June 2017, including impressions, click-through rate (CTR), value-per-click (VPC), average cost-per-click (CPC) and chosen matching types. The dataset contains 34 ad-groups with 627 keywords. In Table 4, we give examples of our dataset with keyword ID, ad-group ID, matching types, and observed keyword performance indices (i.e., impression, CTR, VPC and CPC) over chosen matching types. The unobserved keyword performance indices (over unchosen matching types) are marked with symbolic hyphens "-". Table 5 shows summary statistics of the dataset, including proportions of chosen matching types, and the mean and standard deviation of keyword performance indices over chosen matching types.

**Table 4. Examples of the Dataset**

| Day | Keyword ID | Ad Group ID | Matching Type | Impression | CTR | VPC | CPC |
|---|---|---|---|---|---|---|---|
| 2012/6/13 | keyword-31 | ad-group-13 | exact | - | - | - | - |
| | | | phrase | - | - | - | - |
| | | | broad | 36 | 0.06 | 50 | 0.31 |
| 2012/6/20 | keyword-31 | ad-group-13 | exact | - | - | - | - |
| | | | phrase | - | - | - | - |
| | | | broad | 18 | 0.17 | 18.3 | 0.25 |
| 2012/8/2 | keyword-31 | ad-group-13 | exact | - | - | - | - |
| | | | phrase | - | - | - | - |
| | | | broad | 37 | 0.22 | 25 | 0.26 |
| 2012/9/3 | keyword-31 | ad-group-13 | exact | - | - | - | - |
| | | | phrase | - | - | - | - |
| | | | broad | 49 | 0.16 | 25 | 0.29 |
| 2012/10/20 | keyword-31 | ad-group-13 | exact | - | - | - | - |
| | | | phrase | - | - | - | - |
| | | | broad | 35 | 0.11 | 35 | 0.33 |
| 2013/3/16 | keyword-31 | ad-group-13 | exact | - | - | - | - |
| | | | phrase | - | - | - | - |
| | | | broad | 12 | 0.16 | 24.5 | 0.16 |
| …… | | | | | | | |
| 2016/3/8 | keyword-402 | ad-group-25 | exact | - | - | - | - |
| | | | phrase | 2 | 0.5 | 75 | 0.31 |
| | | | broad | - | - | - | - |
| 2016/11/12 | keyword-402 | ad-group-25 | exact | - | - | - | - |
| | | | phrase | 7 | 0.29 | 39.5 | 0.42 |
| | | | broad | - | - | - | - |
| 2016/11/30 | keyword-402 | ad-group-25 | exact | - | - | - | - |
| | | | phrase | 8 | 0.13 | 53.5 | 0.46 |
| | | | broad | - | - | - | - |
| 2016/12/14 | keyword-402 | ad-group-25 | exact | - | - | - | - |
| | | | phrase | 26 | 0.35 | 10 | 0.33 |
| | | | broad | - | - | - | - |
| 2016/12/17 | keyword-402 | ad-group-25 | exact | - | - | - | - |
| | | | phrase | 25 | 0.24 | 27.3 | 0.28 |
| | | | broad | - | - | - | - |
| 2016/12/21 | keyword-402 | ad-group-25 | exact | - | - | - | - |

| | | | phrase | 52 | 0.13 | 4 | 0.35 |
| | | | broad | - | - | - | - |
| | | | …… | | | | |
| 2013/1/8 | keyword-527 | ad-group-9 | exact | 21 | 0.38 | 39.8 | 0.27 |
| | | | phrase | - | - | - | - |
| | | | broad | - | - | - | - |
| 2013/5/20 | keyword-527 | ad-group-9 | exact | 21 | 0.14 | 6.67 | 0.12 |
| | | | phrase | - | - | - | - |
| | | | broad | - | - | - | - |
| 2013/6/24 | keyword-527 | ad-group-9 | exact | 8 | 0.5 | 14.8 | 0.15 |
| | | | phrase | - | - | - | - |
| | | | broad | - | - | - | - |
| 2013/9/10 | keyword-527 | ad-group-9 | exact | 11 | 0.09 | 100 | 0.15 |
| | | | phrase | - | - | - | - |
| | | | broad | - | - | - | - |
| 2014/1/11 | keyword-527 | ad-group-9 | exact | 8 | 0.25 | 95 | 0.14 |
| | | | phrase | - | - | - | - |
| | | | broad | - | - | - | - |
| 2015/1/10 | keyword-527 | ad-group-9 | exact | 8 | 0.38 | 64.5 | 0.17 |
| | | | phrase | - | - | - | - |
| | | | broad | - | - | - | - |
| | | | …… | | | | |

**Table 5. Summary Statistics of the Dataset**

| Matching Type | Proportion（%） | Impression | | CTR | | VPC | | CPC | |
| --- | --- | --- | --- | --- | --- | --- | --- | --- | --- |
| \ | \ | Mean | SD | Mean | SD | Mean | SD | Mean | SD |
| Exact | 32.31 | 87.06 | 243.12 | 0.45 | 0.24 | | | | |
| Phrase | 15.98 | 83.25 | 175.33 | 0.29 | 0.29 | 35.59 | 287.14 | 0.23 | 0.20 |
| Broad | 51.71 | 189.60 | 226.00 | 0.42 | 0.22 | | | | |

### 4.2 Experimental Setup

In the stage of data distribution estimation, we make inference about keyword performance indices (i.e., impression and click-through rate) by using the statistical tool WinBUGS to implement Bayesian models with the MCMC methodology (Spiegelhalter et al., 2003). We run each process for 50,000 iterations through referring to the reasonable range suggested by related research (e.g., Carrigan et al., 2007; Voleti et al., 2015).

In the stage of stochastic keyword targeting optimization, experiments are coded and run in Python 3.9.7. We set the confidence level (i.e., $\alpha$) for the chance constraint of the budget as 0.95, which is a widely used default number in stochastic programming (Shapiro et al., 2021). In the dataset, the total cost of all keywords is 1175.74. In the following experiments, we gradually increase the budget setting from 100 to 1000 by step of 100 to evaluation the performance of our keyword targeting strategy at different budget levels.

### 4.3 Experimental Results

*4.3.1 Performance Comparisons*

We compare BB-KSM with seven baselines with respect to the expected profit and the profit-cost distribution. The first baseline (BASE1-Past) selects keywords based on the used frequency of keywords in past SSA campaigns by the advertiser. There is limited research on keyword targeting optimization and no suitable approach can be directly compared with ours. For comparison, we derive three baselines from the literature on keyword selection. The second baseline (BASE2-PrefixOrder) derived from (Rusmevichientong and Williamson, 2006) adaptively selects keywords based on a prefix ordering (i.e., sorting keywords in descending order of profit-to-cost ratio) until the expected cost is close to the budget. The third baseline (BASE3-Competitiveness) derived from (Zhang et al., 2014) selects keywords with goals of maximizing the advertiser's profit and minimizing the keyword competitiveness under the budget constraint. The index of "impression confidence based on competitiveness" in Zhang et al. (2014) is defined as $c = 1 - 1/(1 + e^{-\tau d})$ in the baseline, where $\tau > 0$ is a coefficient and $d$ is keyword impression. The fourth baseline (BASE4-SharpeRatio) derived from (Symitsi et al., 2022) treats keywords as risky stocks and applies mean-variance optimization to select keywords in the optimal risky portfolio with the highest Sharpe ratio. In order to validate the effectiveness of our data distribution estimation approach, we develop the fifth baseline (BASE5-SelectNomatch), which uses the stochastic keyword targeting model derived from Section 3.2, with no data distribution estimation process in Section 3.1. As far as we knew, there is no reported research that recommends a specific keyword matching strategy in the literature. For comparison, we designed two baselines related to keyword matching. The sixth baseline (BASE6-RandMatch) selects keywords in a descending order of keyword profit, and then randomly chooses the matching type for each keyword based on the partially observed keyword performance indices, under the budget constraint. The seventh baseline (BASE7-OptMatch) selects keywords in a descending order of keyword profit, and then chooses the optimal matching type for each keyword based on the complete information of keyword performance indices over the three matching types, under the budget constraint. Note that the 1st-5th baselines deal with keyword selection based on the partially observed keyword performance indices, ignoring keyword matching optimization, while the 6th and 7th baselines are about keyword matching.

Figure 1 shows the expected profit obtained by BB-KSM and seven baselines at various budget levels. Figure 2 illustrates the profit and cost of selected keywords by BB-KSM and seven baselines with the budget constraint of 500.

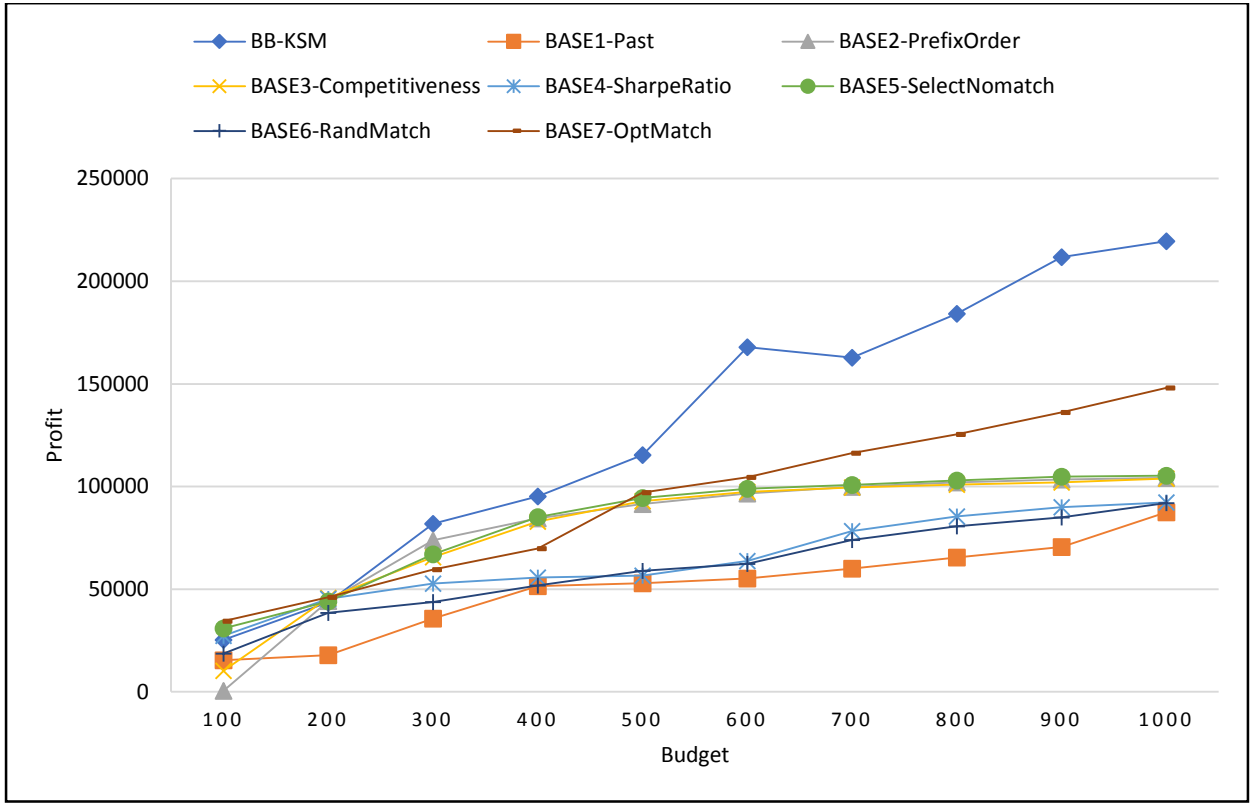

**Figure 1. The Expected Profit Obtained by BB-KSM and Seven Baselines at Various Budget Levels**

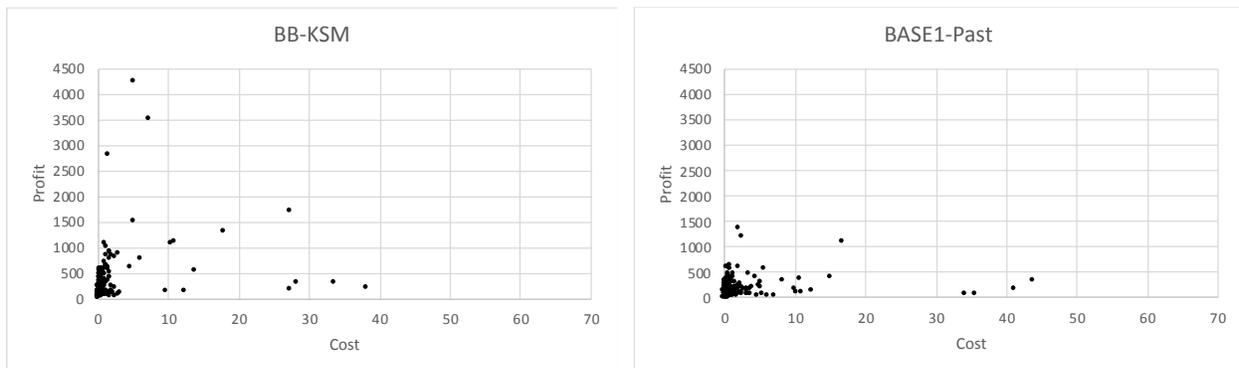

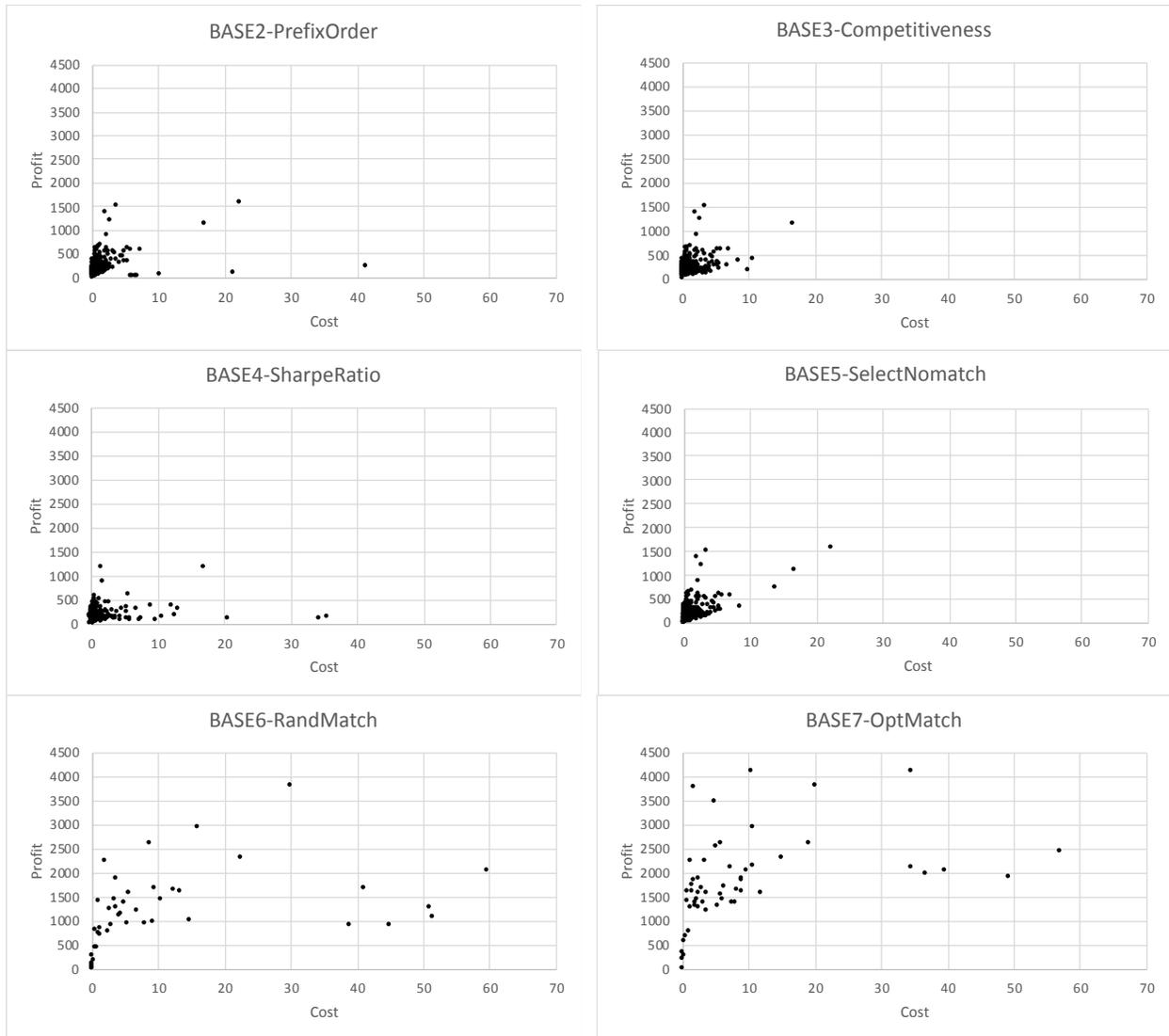

**Figure 2. The Profit and Cost of Keywords Selected by BB-KSM and Seven Baselines**

From Figures 1 and 2, we can observe the following results. First of all, overall BB-KSM outperforms seven baselines in terms of profit. More specifically, BB-KSM performs less well in cases with a small amount of budget; however, when the budget increases, BB-KSM gradually obtains a clear advantage in terms of profit growth, compared with seven baselines. The larger budget makes the keyword targeting decision complicated with more keyword-matching combinations. BB-KSM can find high-profit yet low-cost keyword-matching combinations in complicated environments by searching the keyword targeting space for the global optimum, based on the complete information of keyword performance indices over the three matching types.

Second, compared with BASE5-SelectNomatch, the superiority of BB-KSM proves the effectiveness of the proposed data distribution estimation approach in Section 3.1. That is, our data distribution estimation approach can produce accurate estimations for keyword performance indices over the three matching types, which can substantially help enrich keyword portfolios and increase the profit.

Third, among the five baselines of keyword selection, BASE5-SelectNomatch performs slightly better than the other four baselines (i.e., BASE1-Greedy, BASE2-PrefixOrder, BASE3-Competitiveness and BASE4-SharpeRatio). This result could be attributed to the superiority of using our stochastic optimization model which takes the uncertainty of SSA market into account. That is, our branch-and-bound algorithm combined with stochastic simulation can explore every possibility in the solution space of keyword targeting. Moreover, BASE1-Greedy performs the worst, possibly due to the fact that advertisers usually have limited knowledge and experience on advertising decisions (Yang et al., 2012). In the meanwhile, BASE4-SharpeRatio obtains the second lowest profit. This can be explained as follows. BASE4-SharpRatio assumes that an increase in keyword popularity (i.e., impressions) is associated with an increase in profit, and thus selects keywords based on the mean and variance of the growth rate in keyword popularity. However, there is a complex mechanism connecting keyword popularity and monetization in SSA. That is, impressions do not lead to profits in a simple and direct manner.

Fourth, among the two baselines of keyword matching (i.e., BASE6-RandMatch and BASE7-OptMatch), BASE7-OptMatch performs better than BASE6-RandMatch. This is also owing to the proposed data distribution estimation approach which provides the complete information of performance indices.

*4.3.2 The Number of Selected Keywords*

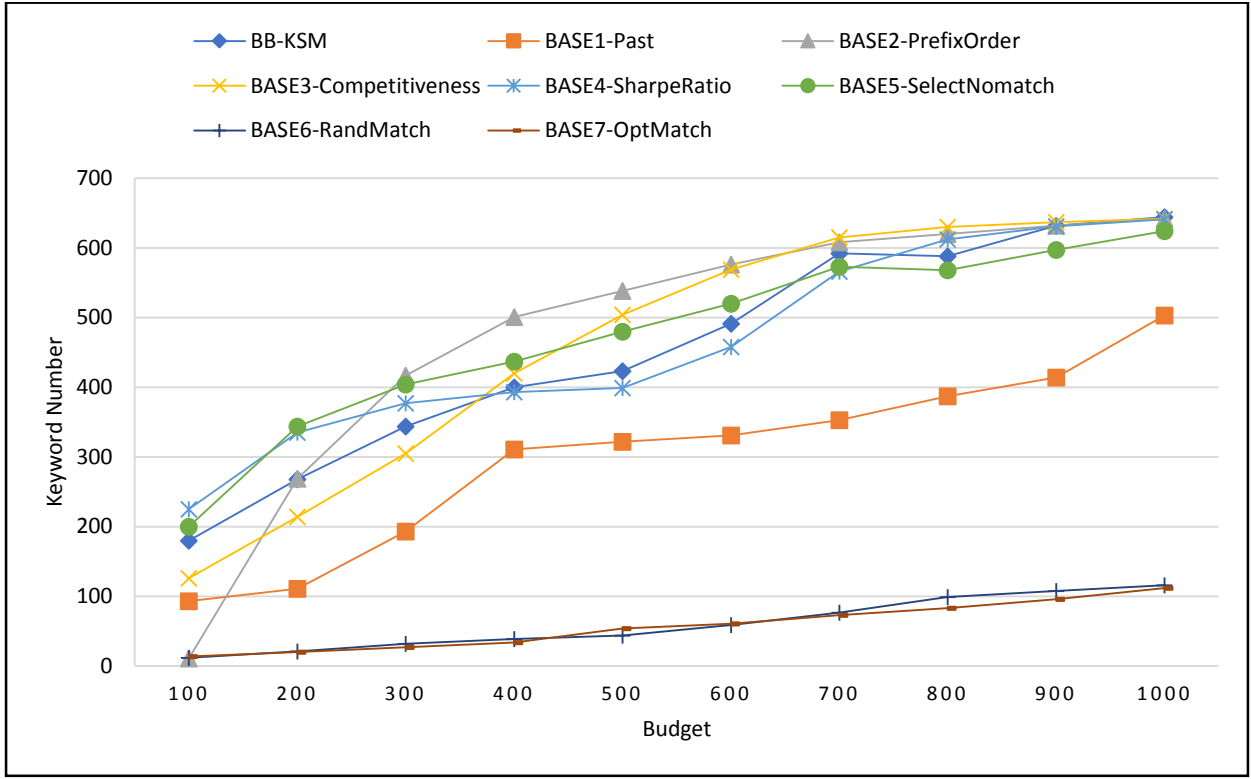

**Figure 3. The Number of Keywords Selected by BB-KSM and Seven Baselines**

The number of keywords selected by our keyword targeting strategy (BB-KSM) and seven baselines at various budget levels are shown in Figure 3.

From Figure 3, we can see that the number of selected keywords by BB-KSM and seven baselines show a similar upward trend with the increase of the budget. However, there are several exceptions where the budget increases but the number of selected keywords decreases. This is because there exists some high-profit and high-cost keywords (i.e., popular keywords) in SSA. Popular keywords may make advertisers run out of the budget too quickly and miss better opportunities to display their advertisements, which may hurt their profits. In addition, selecting more keywords does not necessarily produce a higher profit.

*4.3.3 Keyword Matching Types*

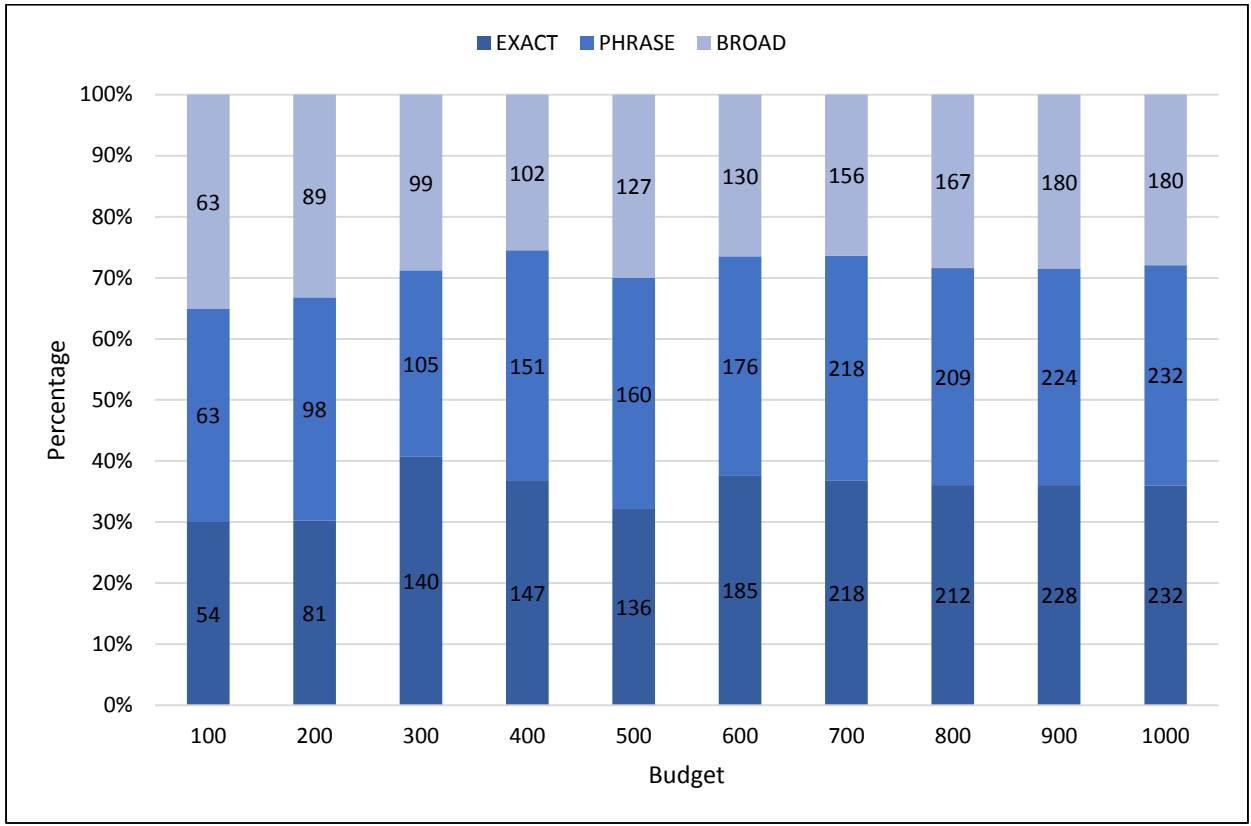

**Figure 4. The Percentage of Keyword Matching Types**

Figure 4 shows the percentage of matching types for selected keywords by BB-KSM at various budget levels. From Figure 4, we can find that exact match and phrase match take up considerable proportions in the optimal keyword targeting solution. This result is not well aligned with the industrial consensus that broad match is much popular than exact match and phrase match: 56% of clicks are from broad match, compared to only 33% from exact match and 11% from phrase match on Google; meanwhile, 70% from broad match, 20% from exact match and 10% from phrase match on Bing (Amaldoss et al., 2016). Our result warns advertisers against choosing broach match by default in SSA. Instead, advertisers should be cautious to determine matching types for selected keywords.

On one hand, broad match maximizes the opportunity for advertisers to reach potential consumers, and in the meanwhile, lead to higher monetization of traffics for search engines (Gupta et al., 2009). On the other hand, it might waste a large amount of advertising budget paying for irrelevant traffics that do not convert (WordStream, 2021), push advertisers into a more intense competition through the wider targeting (Levin and Milgrom, 2010), and thus harm their earnings (Amaldoss et al., 2016; Yang et al., 2021a).

Compared with a single matching type, mixed matching types help balance the tradeoff between reaching the wide target population and avoiding the unnecessary spending on irrelevant clicks. BB-KSM can choose a suitable matching type for each keyword based on historical keyword performance indices, rather than set a one-size-fits-all matching option for all keywords.

## 5. Conclusions

In this paper, we construct a data distribution estimation model and apply the Markov Chain Monte Carlo method to make inference about unobserved performance indices (i.e., impression and click-through rate) over the three matching types. Based on the estimation results, we propose a stochastic keyword targeting model to maximize the expected profit under the chance constraint of the budget, and develop a BB-KSM solution to solve our keyword targeting model. Using a realworld dataset collected from field reports and logs of past SSA campaigns, we conduct a series of experiments to validate the effectiveness of our keyword targeting strategy. Experimental results illustrate that our keyword targeting strategy outperforms seven baselines in terms of the expected profit and our data distribution estimation approach can effectively address the problem of incomplete performance indices over keyword matching types.

### 5.1 Theoretical Contributions

This research makes important contributions in the following aspects. From an academic perspective, to the best of our knowledge, this is the first research effort on the keyword targeting problem in SSA. Previous literatures have studied keyword selection (e.g., Rusmevichientong and Williamson, 2006; Kiritchenko and Jiline, 2008; Zhang et al., 2014; Desai et al., 2014; Lu and Yang, 2017) and keyword matching (e.g., Radlinski et al., 2008; Singh and Roychowdhury, 2008; Even Dar et al., 2009; Gupta et al., 2009; Grbovic et al. 2016; Amaldoss et al. 2016) separately. This study also adds to the advertising optimization literature by conducting the joint optimization of keyword selection and keyword matching in SSA in the framework of stochastic optimization under the chance constraint of the budget. From the methodology perspective, we construct a stochastic keyword targeting model and develop a corresponding branch-and-bound algorithm. This provides a feasible way for advertisers to make keyword targeting decisions in practice. Moreover, we propose a data distribution estimation approach using a fully Bayesian model to address the problem of unobserved keyword performance indices over keyword matching types. Last but not the least, this research enhances the understanding of keyword decisions and is promising to be adapted to other keyword-based advertising forms, e.g., social media advertising and native advertising (Yang and Gao, 2021).

**5.2 Practical Implications**

The results from this research suggest critical managerial insights into keyword decisions for advertisers in SSA. First, keyword targeting is a vitally important advertising decision in SSA. Especially, when the SSA market becomes more complicated, advertisers with a larger number of keywords and keyword combinations should pay more attention on keyword targeting. Second, various factors (e.g., selected keywords, matching types, the control of uncertainty and the budget) have influences on the performance of keyword targeting decisions. Advertisers need to comprehensively evaluate and utilize these factors to get an optimal keyword targeting decision. Third, in the optimal keyword targeting solutions, exact and phrase matching types take up a considerable proportion. This is not well aligned with the common knowledge that broad match is the most popular option (Amaldoss et al., 2016). This reminds advertisers to explore keyword targeting strategies with mixed matching types.

**5.3 Future Research**

In the future, we seek to explore dynamic keyword targeting strategies in SSA, that is, continuously adjusting keyword targeting based on immediate market responses. Moreover, joint optimization of various advertising decisions such as budget planning and keyword portfolios is an interesting perspective to explore in the SSA context. Furthermore, we also plan to adapt our keyword targeting model in other online advertising forms.

## Acknowledgements

We are thankful to the editor and anonymous reviewers who provided valuable suggestions that led to a considerable improvement in the organization and presentation of this manuscript. This work is partially supported by the (NSFC National Natural Science Foundation of China) grants (72171093, 71672067).

# References

Amaldoss, W., Jerath, K., & Sayedi, A. (2016). Keyword management costs and "broad match" in sponsored search advertising. *Marketing Science*, *35*(2), 259-274.

Arroyo-Cañada FJ. and Gil-Lafuente, J. (2019). A fuzzy asymmetric TOPSIS model for optimizing investment in online advertising campaigns. *Operational Research International Journal*, 19(3), 701-716.

Avadhanula, V., Colini Baldeschi, R., Leonardi, S., Sankararaman, K. A., & Schrijvers, O. (2021, April). Stochastic bandits for multi-platform budget optimization in online advertising. In *Proceedings of the Web Conference* 2021 (pp. 2805-2817).

Carrigan, G., Barnett, A., Dobson, A., & Mishra, G. (2007). Compensating for missing data from longitudinal studies using WinBUGS. *Journal of Statistical Software*, *19*(7), 1-17.

Chen, M. H., Shao, Q. M., & Ibrahim, J. G. (2000). *Monte Carlo Methods in Bayesian Computation*. New York: Spinger-Verlag.

Cholette, S.; Özlük, Ö.; and Parlar, M. (2012). Optimal keyword bids in search-based advertising with stochastic advertisement positions. *Journal of Optimization Theory and Applications*, 152(1), 225-244.

Desai, P. S., Shin, W., & Staelin, R. (2014). The company that you keep: When to buy a competitor's keyword. *Marketing Science*, *33*(4), 485-508.

Du, X., Su, M., Zhang, X., & Zheng, X. (2017). Bidding for multiple keywords in sponsored search advertising: Keyword categories and match types. *Information Systems Research*, 28(4), 711-722.

Even Dar, E., Mirrokni, V. S., Muthukrishnan, S., Mansour, Y., & Nadav, U. (2009, April). Bid optimization for broad match ad auctions. In *Proceedings of the 18th International Conference on World Wide Web* (pp. 231-240).

Gamerman, D., & Lopes, H. F. (2006). *Markov chain Monte Carlo: Stochastic simulation for Bayesian inference*. Boca Raton, FL: Chapman & Hall/CRC.

Ghose, A., & Yang, S. (2009). An empirical analysis of search engine advertising: Sponsored search in electronic markets. *Management Science*, 55(10), 1605-1622.

Grbovic, M., Djuric, N., Radosavljevic, V., Silvestri, F., Baeza-Yates, R., Feng, A., Ordentlich, E., Yang, L. & Owens, G. (2016, July). Scalable semantic matching of queries to ads in sponsored search advertising. In *Proceedings of the 39th International ACM SIGIR Conference on Research and Development in Information Retrieval* (pp. 375-384).

Gupta, S., Bilenko, M., & Richardson, M. (2009, June). Catching the drift: learning broad matches from clickthrough data. In *Proceedings of the 15th ACM SIGKDD International Conference on Knowledge Discovery and Data Mining* (pp. 1165-1174).

Hao, X., Peng, Z., Ma, Y., Wang, G., Jin, J., Hao, J., et al. (2020, November). Dynamic knapsack optimization towards efficient multi-channel sequential advertising. In *International Conference on Machine Learning* (pp. 4060-4070).

Huang, H., & Kauffman, R. J. (2011). On the design of sponsored keyword advertising slot auctions: An analysis of a generalized second-price auction approach. *Electronic Commerce Research and Applications*, 10(2), 194-202.

Interactive Advertising Bureau. *2021 Marketplace Outlook Survey Results*. Available at https://www.iab.com/wp-content/uploads/2020/12/2021-IAB-Marketplace-Outlook-Dec-2020-FINAL.pdf (accessed on January 28, 2022).

Jeziorski, P., & Moorthy, S. (2018). Advertiser prominence effects in search advertising. *Management Science*, 64(3), 1365-1383.

Jones, R., Rey, B., Madani, O., & Greiner, W. (2006, May). Generating query substitutions. In *Proceedings of the 15th International Conference on World Wide Web* (pp. 387-396).

Kim, A. J., Jang, S., & Shin, H. S. (2021). How should retail advertisers manage multiple keywords in paid search advertising?. *Journal of Business Research*, 130, 539-551.

Kiritchenko, S., & Jiline, M. (2008, September). Keyword optimization in sponsored search via feature selection. In *Proceedings of the Workshop on New Challenges for Feature Selection in Data Mining and Knowledge Discovery* (pp. 122-134).

Kosuch, S., and Lisser, A. (2010). Upper bounds for the 0–1 stochastic knapsack problem and a B&B algorithm. *Annals of Operations Research*, 176(1), 77–93.

Küçükaydin, H., Selçuk, B., & Özlük, Ö. (2020). Optimal keyword bidding in search-based advertising with budget constraint and stochastic ad position. *Journal of the Operational Research Society*, 71(4), 566-578.

Levin, J., & Milgrom, P. (2010). Online advertising: Heterogeneity and conflation in market design. *American Economic Review*, 100(2), 603-607.

Li, H., & Yang, Y. (2020). Optimal keywords grouping in sponsored search advertising under uncertain environments. *International Journal of Electronic Commerce*, 24(1), 107-129.

Li, H., Kannan, P. K., Viswanathan, S., & Pani, A. (2016). Attribution strategies and return on keyword investment in paid search advertising. *Marketing Science*, 35(6), 831-848.

Lian, Y., Chen, Z., Pei, X., Li, S., Wang, Y., Qiu, Y., et al. (2021, April). Optimizing AD Pruning of Sponsored Search with Reinforcement Learning. In *Companion Proceedings of the Web Conference 2021* (pp. 123-127).

Lo, S. K., Hsieh, A. Y., & Chiu, Y. P. (2014). Keyword advertising is not what you think: Clicking and eye movement behaviors on keyword advertising. *Electronic Commerce Research and Applications*, 13(4), 221-228.

Lobo, M. S., Vandenberghe, L., Boyd, S. & Lebret, H. (1998). Applications of second-order cone programming. *Linear Algebra and Its Applications*, 284(1-3), 193-228.

Lu, S., & Yang, S. (2017). Investigating the spillover effect of keyword market entry in sponsored search advertising. *Marketing Science*, 36(6), 976-998.

Mahdian, M., & Wang, G. (2009, September). Clustering-based bidding languages for sponsored search. In *European Symposium on Algorithms* (pp. 167-178), Springer, Berlin, Heidelberg.

Nagpal, M., & Petersen, J. A. (2020). Keyword Selection Strategies in Search Engine Optimization: How Relevant is Relevance?. *Journal of Retailing*, 97(4), 746-763.

Nie, H., Yang, Y., & Zeng, D. (2019). Keyword generation for sponsored search advertising: Balancing coverage and relevance. *IEEE Intelligent Systems*, 34(5), 14-24.

Nuara, A., Trovò, F., Gatti, N., & Restelli, M. (2022). Online joint bid/daily budget optimization of internet advertising campaigns. *Artificial Intelligence*, 305, 103663.

Prekopa, A. *Stochastic Programming*. Dordrecht, the Netherlands: Kluwer Academic, 1995.

Qiao, D., Zhang, J., Wei, Q., & Chen, G. (2017). Finding competitive keywords from query logs to enhance search engine advertising. *Information & Management*, 54(4), 531-543.

Radlinski, F., Broder, A., Ciccolo, P., Gabrilovich, E., Josifovski, V., & Riedel, L. (2008, July). Optimizing relevance and revenue in ad search: a query substitution approach. In *Proceedings of the 31st Annual International ACM SIGIR Conference on Research and Development in Information Retrieval* (pp. 403-410).

Ramaboa, K. K., & Fish, P. (2018). Keyword length and matching types as indicators of search intent in sponsored search. *Information Processing & Management*, *54*(2), 175-183.

Rusmevichientong, P., & Williamson, D. P. (2006, June). An adaptive algorithm for selecting profitable keywords for search-based advertising services. In *Proceedings of the 7th ACM Conference on Electronic Commerce* (pp. 260-269).

Scholz, M., Brenner, C., & Hinz, O. (2019). AKEGIS: automatic keyword generation for sponsored search advertising in online retailing. *Decision Support Systems*, 119, 96-106.

Schultz, C. D. (2020). The impact of ad positioning in search engine advertising: a multifaceted decision problem. *Electronic Commerce Research*, 20(4), 945-968.

Shapiro, A., Dentcheva, D., & Ruszczynski, A. (2021). *Lectures on stochastic programming: modeling and theory*. Society for Industrial and Applied Mathematics.

Singh, S. K., & Roychowdhury, V. P. (2008) To broad-match or not to broad-match: An auctioneer's dilemma. In *Proceedings of the fourth Workshop on Ad Auctions, Conference on Electronic Commerce (EC'08)* (pp. 1-33).

Song, Z., Chen, J., Zhou, H., & Li, L. (2021, March). Triangular Bidword Generation for Sponsored Search Auction. In *Proceedings of the 14th ACM International Conference on Web Search and Data Mining* (pp. 707-715).

Spiegelhalter, D. J., Thomas, A., Best, N., & Lunn, D. (2003). *WinBUGS version 1.4 user manual*. MRC Biostatistics Unit, Cambridge. Available at https://www.mrc-bsu.cam.ac.uk/wp-content/uploads/manual14.pdf (accessed on January 28, 2022).

Statista. *Search Advertising*. Available at https://www.statista.com/outlook/219/100/search-advertising/worldwide (accessed on January 28, 2022).

Symitsi, E., Markellos, R. N., & Mantrala, M. K. (2022). Keyword portfolio optimization in paid search advertising. *European Journal of Operational Research*, 303(2),767-778.

Szymanski, G., & Lininski, P. (2018, September). Model of the effectiveness of Google Adwords advertising activities. In *2018 IEEE 13th International Scientific and Technical Conference on Computer Sciences and Information Technologies (CSIT)* (Vol. 2, pp. 98-101).

Voleti, S., Kopalle, P. K., & Ghosh, P. (2015). An interproduct competition model incorporating branding hierarchy and product similarities using store-level data. *Management Science*, 61(11), 2720-2738.

Vragov, R., Di Shang, R., Smith, V., & Porter, D. (2019). Let's play the search game: Strategic and behavioral properties of sponsored search auction mechanisms. *Electronic Commerce Research and Applications*, 33, 100809.

Wächter, A. & Biegler, L. T. (2006). On the implementation of an interior-point filter line-search algorithm for large-scale nonlinear programming. *Mathematical Programming,* 106(1), 25-57.

WordStream. *Google Ads Match Types: How Do Keyword Match Types Work in Google?* Available at https://www.wordstream.com/keyword-match-types (accessed on January 28, 2022).

Yang, C., & Xiong, Y. (2020). Nonparametric advertising budget allocation with inventory constraint. *European Journal of Operational Research*, 285(2), 631-641.

Yang, S., Pancras, J., & Song, Y. A. (2021a). Broad or exact? Search Ad matching decisions with keyword specificity and position. *Decision Support Systems*, 143, 113491.

Yang, W., Xiao, B., & Wu, L. (2020). Learning and pricing models for repeated generalized second-price auction in search advertising. *European Journal of Operational Research*, 282(2), 696-711.

Yang, Y. and Zhai, P. (2022). Click-through rate prediction in online advertising: A literature review. *Information Processing & Management*, 59(2), 102853.

Yang, Y., & Gao, T. L. (2021). The path to people's responses to native advertising in social media: A perspective of self-presentational desire. *Information & Management*, 58(3), 103441.

Yang, Y., Feng, B., & Zeng, D. (2021c). A Generalized Vidale-Wolfe Response Model with Flexible Ad Elasticity and Word-of-Mouth. *IEEE Intelligent Systems,* 36(5), 69-79.

Yang, Y., Feng, B., Salminen, J., & Jansen, B. J. (2021b). Optimal advertising for a generalized Vidale–Wolfe response model. *Electronic Commerce Research*, forthcoming. https://doi.org/10.1007/s10660-021-09468-x.

Yang, Y., Jansen, B. J., Yang, Y., Guo, X., & Zeng, D. (2019). Keyword optimization in sponsored search advertising: A multilevel computational framework. *IEEE Intelligent Systems*, 34(1), 32-42.

Yang, Y., Li, X., Zeng, D., & Jansen, B. J., (2018). Aggregate Effects of Advertising Decisions: A Complex Systems Look at Search Engine Advertising via an Experimental Study. *Internet Research*, 28(4), 1079-1102.

Yang, Y., Qin, R., Jansen, B. J., Zhang, J., & Zeng, D. (2014). Budget planning for coupled campaigns in sponsored search auctions. *International Journal of Electronic Commerce*, 18(3), 39-66.

Yang, Y., Yang, Y. C., Jansen, B. J. and Lalmas, M. (2017). Computational Advertising: A Paradigm Shift for Advertising and Marketing?. *IEEE Intelligent Systems,* 32(3), 3-6.

Yang, Y., Zeng, D., Yang, Y., and Zhang, J. (2015). Optimal budget allocation across search advertising markets. *INFORMS Journal on Computing*, 27(2), 285-300.

Yang, Y., Zhang, J., Qin, R., Li, J., Liu, B., and Liu, Z. (2013). Budget strategy in uncertain environments of search auctions: A preliminary investigation. *IEEE Transactions on Services Computing*, 6(2), 168-176.

Yang, Y., Zhang, J., Qin, R., Li, J., Wang, F. Y., & Qi, W. (2012). A budget optimization framework for search advertisements across markets. *IEEE Transactions on Systems, Man, and Cybernetics-Part A: Systems and Humans*, 42(5), 1141-1151.

Yang, Y., Zhao, K., Zeng, D., and Jansen, J.B. (2022). Time-varying Effects of Search Engine Advertising on Sales--An Empirical Investigation in E-Commerce. *Decision Support Systems*, forthcoming. https://doi.org/10.1016/j.dss.2022.113843.

Zhang, J., Zhang, J., & Chen, G. (2021). A semantic transfer approach to keyword suggestion for search engine advertising. *Electronic Commerce Research*, forthcoming. https://doi.org/10.1007/s10660-021-09496-7.

Zhang, W., Zhang, Y., Gao, B., Yu, Y., Yuan, X., & Liu, T. Y. (2012, August). Joint optimization of bid and budget allocation in sponsored search. In *Proceedings of the 18th ACM SIGKDD International Conference on Knowledge Discovery and Data Mining* (pp. 1177-1185).

Zhang, Y., Zhang, W., Gao, B., Yuan, X., & Liu, T. Y. (2014). Bid keyword suggestion in sponsored search based on competitiveness and relevance. *Information Processing & Management*, 50(4), 508-523.